\documentclass{jfm}
\usepackage{multirow}
\usepackage{graphicx}
\usepackage{natbib}
\usepackage{epstopdf}
\usepackage[normalem]{ulem}
\ifCUPmtlplainloaded \else
  \checkfont{eurm10}
  \iffontfound
    \IfFileExists{upmath.sty}
      {\typeout{^^JFound AMS Euler Roman fonts on the system,
                   using the 'upmath' package.^^J}%
       \usepackage{upmath}}
      {\typeout{^^JFound AMS Euler Roman fonts on the system, but you
                   dont seem to have the}%
       \typeout{'upmath' package installed. JFM.cls can take advantage
                 of these fonts,^^Jif you use 'upmath' package.^^J}%
       \providecommand\upi{\pi}%
      }
  \else
    \providecommand\upi{\pi}%
  \fi
\fi

\ifCUPmtlplainloaded \else
  \checkfont{msam10}
  \iffontfound
    \IfFileExists{amssymb.sty}
      {\typeout{^^JFound AMS Symbol fonts on the system, using the
                'amssymb' package.^^J}%
       \usepackage{amssymb}%
         \let\leq=\leqslant
         \let\geq=\geqslant
      }{}
  \fi
\fi

\ifCUPmtlplainloaded \else
  \IfFileExists{amsbsy.sty}
    {\typeout{^^JFound the 'amsbsy' package on the system, using it.^^J}%
     \usepackage{amsbsy}}
    {}
\fi

\newsavebox{\astrutbox}
\sbox{\astrutbox}{\rule[-5pt]{0pt}{20pt}}

\usepackage{subfig}
\usepackage{float}
\usepackage{soul}
\usepackage[usenames]{color}
\usepackage{natbib}
\usepackage{amsfonts}
\usepackage{amsmath}

\title{Role of the basin boundary conditions in gravity wave turbulence}

\author[]{L.~Deike$^{1}$, B.~Miquel$^2$, P.~Guti\'{e}rrez$^3$, T.~Jamin$^{1}$, B.~Semin$^2$, M.~Berhanu$^{1}$, E.~Falcon$^{1}$\thanks{Email address for correspondence: eric.falcon@univ-paris-diderot.fr}, and F.~Bonnefoy$^4$}
\affiliation{$^1$Univ Paris Diderot, Sorbonne Paris Cit\'e, MSC, UMR 7057 CNRS, F-75 013 Paris, France \\[\affilskip]
$^2$Ecole Normale Sup\'erieure, LPS, UMR 8550 CNRS, F-75 205 Paris, France \\[\affilskip]
$^3$CEA-Saclay, Sphynx, DSM, URA 2464 CNRS, F-91 191 Gif-sur-Yvette, France \\[\affilskip]
$^4$Ecole Centrale de Nantes, LHEEA, UMR 6598 CNRS, F-44 321 Nantes, France}

\begin{document}
\maketitle
\begin{abstract}
Gravity wave turbulence is  investigated experimentally in a large wave basin  in which irregular waves are generated unidirectionally. The roles of the basin boundary conditions (absorbing or reflecting) and of the forcing properties are investigated. To that purpose, an absorbing sloping beach opposite the wavemaker can be replaced by a reflecting vertical wall. We observe that the wave field properties depend strongly on these boundary conditions.  A quasi-one-dimensional field of nonlinear waves propagates toward the beach where they are damped whereas a more multidirectional wave field is observed with the wall. In both cases, the wave spectrum scales as a frequency-power law with an exponent that increases continuously with the forcing amplitude up to a value close to $-4$. The physical mechanisms involved most likely differ with the boundary condition used, but cannot be easily discriminated with only temporal measurements.  We also studied freely decaying gravity wave turbulence in the closed basin. No self-similar decay of the spectrum is observed, whereas its Fourier modes decay first as a time power law due to nonlinear mechanisms, and then exponentially due to linear viscous damping. We estimate the linear, nonlinear and dissipative time scales to test the time scale separation that highlights the important role of a large scale Fourier mode. By estimation of the mean energy flux from the initial decay of wave energy, the Kolmogorov-Zakharov constant of the weak turbulence theory is evaluated and found to be compatible with a recently obtained theoretical value.
\end{abstract}

\section{Introduction}
The oceanic surface is characterized by the propagation of gravity waves generated by the interaction between the wind and a liquid surface~\citep{Janssen2004}. As the wind distribution over the ocean is inhomogeneous and erratic, forecasting of  sea states is a complex problem. Once generated, the wave field evolves  due to interactions between nonlinear waves, wave dispersion, and dissipation. In particular, when wave amplitudes are high enough, a regime of wave turbulence can be observed, in which the wave field displays a continuous wave spectrum from large to small scales, typically from 100 m to 10 m  \citep{Hwang00}. Wave turbulence theory in its weakly nonlinear limit, also called weak turbulence, yields a theoretical framework to study wave turbulence regimes. This theory provides, for idealistic conditions, an analytical derivation of the spectrum of waves in a turbulent regime in almost all fields of physics involving waves \citep{Zakharovbook,Newell2011,Nazarenkobook}. This theory consists of a weakly nonlinear development of a random field of waves propagating without dissipation in an infinite system. For gravity waves, the spectrum of wave amplitude $S_{\eta}(f)$ is predicted to scale as a frequency-power law of $f^{-4}$, and reads \citep{Zakharov1967,Zakharov82}
\begin{equation}
S_{\eta}(\omega) = C\epsilon^{1/3}g\omega^{-4}\ {\rm ,}
\label{gravsptheo}
\end{equation}
where $\epsilon$ is the mean energy flux, $g$ the acceleration due to gravity, $\omega=2\upi f$, and $C$ the non dimensional Kolmogorov-Zakharov constant. As the hypotheses used are too restrictive to be all verified experimentally, it seems unlikely that wave turbulence theory can alone explain the dynamics of the ocean surface. Nevertheless, this theory can give insights into the mechanisms at play. {\it In situ} observations provide ocean surface measurements for different wind forcing conditions. This has led to several phenomenological descriptions of the wave spectrum that depend on numerous parameters such as duration of wind blowing, wind directionality, fetch length, stage of storm growth and decay, existence of a swell, {\it etc.} \citep{Ochi1998}. As a consequence, {\it in situ} measurements of the wave spectrum varies considerably according to the conditions and locations of observations \citep{Liu89,Banner90}. However, certain measurements of the spectrum are compatible with a $f^{-4}$ scaling \citep{Donelan85,Forristall81,Kahma81,Toba73,Hwang00} thus suggesting a possible agreement with weak turbulence theory at large scale (wavelengths $10 \lesssim \lambda \lesssim 100$ m). At smaller scales ($\lambda < 10$ m), a transition to a steeper spectrum in $f^{-5}$ has been reported \citep{Long2007,Forristall81,RomeroJPO2010}, known as a ``saturation range spectrum'' or the Phillips' spectrum \citep{Phillips58,Kitaigorodskii83}. Occurrence of this steeper spectrum may be caused by wave breakings dissipating all the injected power and by gravity-capillary wave conversion whereas the location of transition scale depends on the wind intensity. But as meteorological conditions are by nature variable and precise measurements  of the ocean surface are difficult, description of this transition between these two kinds of spectra remains an open question. Moreover, the frequency power-law exponent of the spectrum has  been found to depend continuously on the wave steepness \citep{Huang1981}. Laboratory experiments in large wave basins, in which the dynamics of gravity waves produced by wavemaker are studied in well-controlled conditions could thus be useful to better understand out-of-equilibrium spectra of wave elevation in absence of wind forcing.

\begin{table}
\centering
\begin{tabular}{l c c c c c c c}
 & Paris1 & Paris2 & Paris3 & ParisA & Hull & Nantes \\
\hline 
 Basin size L or $L\times l$ (m) & 0.2 &  0.2 & $0.5\times 0.4$ & $1.8\times 0.6 $ & $12\times 6$ & $15\times 10$ \\
Geometry & circular & circular & rect. & rect. & rect. & rect.\\
\hline
Forcing mechanism & pistons & horizontal &  pistons & pistons & pistons & pistons\\
Forcing freq. bandwidth (Hz) & 2 - 6 & 1 - 7 & 1 - 4 & 0-1.5/0-4 & 1 - 1.15 & 1 -1.15\\
Max. spectrum freq. $f_m$ (Hz) & 4 & 4 & 3 & 3 & 1.1 & 1.1\\
Forcing wavelength $\lambda_m$ (m) & 0.1 &0.1& 0.2 & 0.2 &1.4 & 1.4\\
Wave steepness $k_m \times \sigma_{\eta}$ & - & 0.01 - 0.1 & - & - & 0.08 - 0.25 & 0.05 - 0.25 \\  
\hline
$L/\lambda_m$ & 2 & 2 & 2 & 9 & 9 & 11 \\
Piston-gauge distance & 0.7$\lambda_m$ & $\lambda_m$  & $\lambda_m$ & - & 4.3$\lambda_m$ & 5.3$\lambda_m$\\
\hline
\hline
Exponent $\alpha$ for an &-7 to -4 & $-4.5 \pm 0.2$ & -7 to -4 & -6 to -5 & -6.2 to -4 & -8 to -3.5\\
 increasing forcing & & & & $-4\pm1$ &  & &
\end{tabular}
\caption{Previous laboratory experiments on stationary gravity wave turbulence.  Wave spectrum $S_{\eta}(f)$ scales as $f^{\alpha}$ with $\alpha$ depending on the forcing amplitude for several experimental conditions. Paris1 \citep{Falcon2007}, Paris2 \citep{Issenmann13}, Paris3 \citep{Herbert2010}, ParisA \citep{Cobelli2011}, Hull \citep{Denissenko2007,NazarenkoJFM2010}, and Nantes (this article). Working fluid: water except in Paris1 (water or mercury).}
\label{tabmanip}
\end{table}

\section{State of the art concerning gravity wave turbulence in the laboratory}

We limit ourselves here to laboratory experiments on gravity wave turbulence forced by vibrating blades with no wind generation. Recently, several well-controlled experiments have been carried out specifically to test wave turbulence theory for gravity waves on the surface of water \citep{Falcon2007,Denissenko2007,Herbert2010,NazarenkoJFM2010,Cobelli2011,Issenmann13}. The main parameters of these experiments are summarized in table \ref{tabmanip} for the purposes of comparison. The wave spectrum is usually inferred from a capacitive or resistive gauge measuring the temporal wave elevation $\eta(t)$ at a given location. It is defined as the square modulus of the Fourier transform of $\eta(t)$ over a duration $T$,
\begin{equation}
S_{\eta}(f)\equiv \frac{1}{2\upi T}\left|\int_0^T \eta(t)e^{i\omega t}dt\right|^2\ {\rm ,}
\end{equation} where $\omega=2\upi f$. At sufficiently high forcing, the spectrum is found to scale as $f^{\alpha}$ within an inertial range corresponding to gravity wave scales (typically from the forcing scales to centimeter). In most of these experiments in table \ref{tabmanip}, $\alpha$ is found to increase with forcing amplitude for all the basin sizes used (ranging from 20 cm to 15 m), and even when using a low viscosity working fluid such as mercury. When the forcing increases, $\alpha$ increases roughly from $-7$, saturating close to $-4$, the value expected theoretically by weak turbulence. However, this dependence on the forcing amplitude is in strong disagreement with theory. If instead of using a spatially localized forcing (vibrating blades), the whole container is horizontally vibrated (spatially extended forcing), $\alpha$ is found to be independent of the wave steepness over a one-decade frequency range \citep{Issenmann13}. This suggests that the previous discrepancy could be related to the inhomogeneity and anisotropy of the localized forcing. However, the inertial range in the horizontally vibrated experiment was too small to be fully confident of this.

Laboratory measurements of gravity wave height in a turbulent regime, resolved in time and 1D space, have  been performed to better resolve the wave field dynamics \citep{NazarenkoJFM2010}. Both the wavenumber and frequency power law spectra are found to be dependent on the wave strength. Subsequently, measurements have been achieved that are fully resolved in time and 3D space \citep{Herbert2010,Cobelli2011}. The spatial and temporal spectrum scalings were also found to be in strong disagreement with predictions. Presence of strongly nonlinear wave propagation (such as bound waves) have been highlighted leading to a deviation from the linear dispersion relation. As a direct consequence, inferring the spatial $k$-spectrum from the temporal $f$-spectrum by using this dispersion relation yields spurious results. Finally, experiments have underlined the influence of the forcing frequency bandwidth \citep{Cobelli2011}. Indeed, for a narrow forcing frequency bandwidth, the dispersion relation is found to stay close to the linear relation with no bound waves, and $k$- and $f$- spectra seem to be compatible with wave turbulence theory. However, the inertial range of the power-law spectrum is less than half a decade, and for a small forcing amplitude range. Note that the probability distribution of random gravity wave elevation, and the role of the forcing directionality have  also been studied in large wave tanks but without discussing the scaling of the spectrum tail \citep{Onorato09}.

Several explanations have been offered for the dependence of the spectrum exponent on forcing amplitude. First, finite size effects could occur. Some wavelengths are quantized in finite size systems, and the resonant nonlinear wave interactions used in the theoretical derivation are replaced by quasi-resonances \citep{Kartashova98,Zakharov05,Lvov06}. Depletion of pure resonances causes the turbulent transfer through the scales to theoretically become slower and the spectrum steeper \citep{Nazarenko2006}. However, by comparing the experimental data in table \ref{tabmanip}, we do not notice significant differences in $\alpha$ that could be ascribed to finite size effects for all values of basin sizes, of the ratio between the typical forcing wavelength to the basin size, and the piston-gauge distance. Second, the presence of strongly nonlinear waves may  explain the discrepancy with the theory. For instance, sharp crested waves, propagating breaking waves, bound waves or vertical splashes generally occur at different scales and could induce an additional dissipation acting at all scales within the inertial range. These singular coherent structures have a broad signature in Fourier space. Indeed, the spectrum of singularities propagating without deformation ($\omega \sim k$) scales theoretically as $f^{-3-D}$, where $0 \leq D < 2$ is the spatial fractal dimensionality of the coherent structure \citep{Connaughton03}. For instance, if sharp-crested structures occur along ridges ($D=1$), then their spectrum scales as $f^{-4}$ \citep{Kuznetsov2004}. Note that this exponent is similar to that computed by weak turbulence theory (where no crested waves are involved). In the same way, when these wave slope divergences are assumed to be isolated peaks or cusps ($D=0$) distributed isotropically and propagating as $\omega=\sqrt{gk}$, the $f^{-5}$ Phillips'~spectrum is found again. Experimentally, it has been shown that intermittency occurs in gravity wave turbulence \citep{Falcon07b,NazarenkoJFM2010}, and is enhanced by coherent structures such as breaking waves  \citep{Falcon10b}. Third, strongly nonlinear waves involved in laboratory experiments may  lead to non-local interactions in $k$-space, dissipation at all scales of the cascade (energy flux not conserved), and no scale separation between linear, nonlinear, and dissipating time scales, unlike weak turbulence hypotheses. Finally, it has been recently reported in different experimental systems of wave turbulence that increasing dissipation leads to a spectrum that departs from weak turbulence prediction \citep{HumbertEPL2013,MiquelPRE2014,Deike2014}. Note that several numerical simulations of gravity wave turbulence validated the weak turbulence derivation  \citep{Onorato02,Pushkarev03,Dyachenko04,Yokoyama04,Lvov06,Korotkevitch08}. Limited inertial range (no larger than one decade), nonlinearity truncation, and artificial numerical dissipation at large scales are the main  obstacles to  further comparisons of simulation and observations of gravity wave turbulence.

Previous laboratory experiments on gravity wave turbulence have been carried out in closed basins whereas oceans are open systems  for even the largest wavelengths. The reflecting boundary condition used in the laboratory significantly changes  the wave field dynamics with respect to the oceanographic situation. Indeed, in laboratory experiments wave mixing is increased, and counter-propagating waves generate strong splashes. 

In this article, we report an investigation of gravity wave turbulence in a large basin using accurate wave probes. We observe a power-law wave spectrum across a frequency-range of almost two decades, one decade in the gravity range and one in the capillary range. Starting with a closed basin, we confirm previous results on gravity wave turbulence, and extend them to a larger inertial range as well as various experimental parameters (see last column of table \ref{tabmanip}). Then, proceeding with the same basin but with an absorbing boundary condition (beach), we observe similar frequency scalings of the wave spectrum to those observed in the closed basin. Although direct observations of the wave field are observed to be very different for the closed or open basin, the frequency spectra are found to depend on the forcing amplitude with the same trend in both cases. We emphasize  that the physical mechanisms leading to these spectra are likely to be different, and in both cases cannot be  described by weak turbulence theory (interaction between weakly nonlinear resonant waves) alone. In Sect. \ref{inter}, intermittency properties of gravity wave turbulence are quantified. The  value of the intermittency coefficient is found to be roughly the same as in the presence of either beach or wall, suggesting the importance of the coherent structures in both cases. Finally, in Sect. \ref{decay}, we study the non-stationary regime of gravity wave turbulence during its free decay. No self-similar decay is observed in the gravity regime (the frequency power-law exponent of the instantaneous spectrum being dependent on time). We also show that the spectrum Fourier mode amplitudes first decay as a time power law of $t^{-1/2 }$ (as found experimentally by \cite{Nazarenkojetp2013} and predicted theoretically for four-wave interaction systems), and then decrease exponentially over time due to viscous damping. The linear, nonlinear, and dissipative time scales are then inferred at all scales of the cascade. The time scale separation is then tested, and the important role of a large scale Fourier mode (near the forcing scale) for gravity wave turbulence in large basins is highlighted. By estimation of the mean energy flux from the initial decay of wave energy, the Kolmogorov-Zakharov constant is experimentally evaluated for the first time, and found to be compatible with a theoretical value estimated by \cite{Zakharov2010}.

\begin{figure}
 \begin{center}
\includegraphics[scale=0.6]{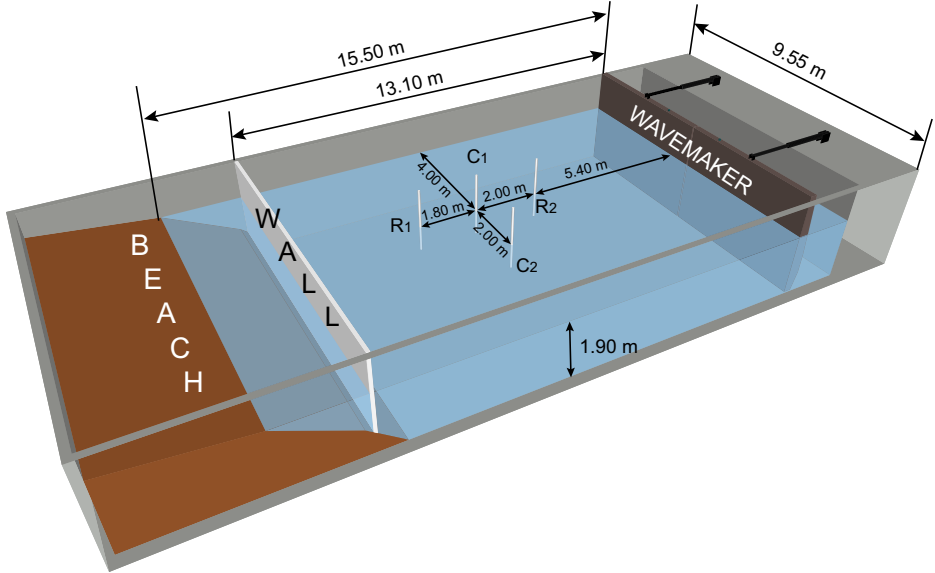} 
\caption{Sketch of the wave basin. The wavemaker is located . The boundary oppos the wavemaker is either a beach or a removable wall. The location of the array of probes is visible: capacitive probes $C_1$ and $C_ 2$, and resistive ones $R_1$ and $R_2$.} 
    \label{flumeschema}
       \end{center}
 \end{figure}

\section{Experimental setup}
\label{setup}
\subsection{Basin and wave generation}
The experiments were performed in a large rectangular wave basin, 15 m $\times$ 10 m, at the Ecole Centrale de Nantes, France. The basin is filled with water with a uniform depth fixed at $1.90$ m. Surface waves are generated by a 10 m wide, rectangular flap wavemaker. The latter is located at one basin width as shown in figure \ref{flumeschema}. This flap is moved by hydraulic cylinders, driven in-phase and controlled by a computer. The wavemaker has a frequency cut-off at 2 Hz due to mechanical parts. A linear variable displacement transducer (LVDT) is fixed on top of the wavemaker  to infer its temporal displacement especially to allow feedback control of the wavemaker position with respect to a prescribed shape spectrum. 

\begin{figure}
 \begin{center}
\includegraphics[scale=0.4]{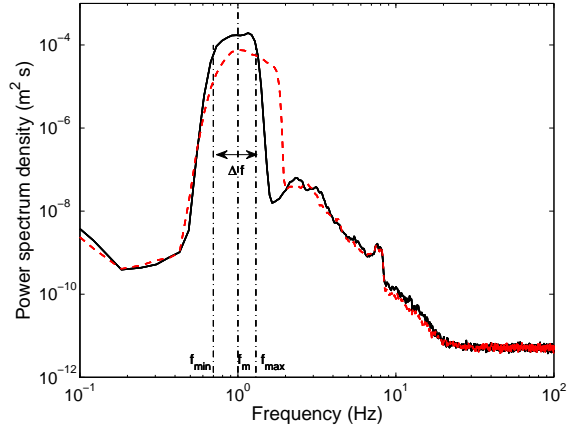} %fig2.eps
\caption{Power spectrum density of wavemaker displacement. Filtered random noise ($-$) and Jonswap ($--$) forcings for a broad frequency bandwidth $\Delta f=f_{max}-f_{min}$. See table \ref{tabnantes} for corresponding forcing parameters.} 
    \label{spectresbatteurs}
       \end{center}
 \end{figure}

The wavemaker generate irregular waves which are randomly distributed in amplitude and in frequency within a certain bandwidth. The wavemaker is driven either by a bandpass filtered random noise (FRN) within a bandwidth $\Delta f$ around a frequency $f_m$  or by a unidirectional Jonswap spectrum (JON - see figure \ref{spectresbatteurs}). The latter has been used in oceanography to model the wave energy in the frequency domain, and based on a parametrization of the wave spectrum measurements in the North Sea \citep[see][]{bookKomen}. In both cases, the forcing parameters are controlled by the frequency bandwidth $\Delta f$ of the spectrum around its maximal value of frequency $f_m$, and by the wavemaker amplitude. Instead of the latter, we will use in the following the value of the rms wave amplitude $\sigma_{\eta}\equiv\sqrt{\overline{ \eta^2(t)}}$ at the gauge locations (temporal average denoted by $\overline{\ \cdot \ }$). Typically,  $f_m \approx 1$ Hz (corresponding to a wavelength $\lambda_m \approx 1.5$ m), $0.3 \leq \Delta f \leq 1.3$ Hz, and $0.5 < \sigma_{\eta} < 7$ cm. Subsequently, frequency bandwidth will be considered in two typical ranges: narrow banded for $\Delta f < 0.5$\,Hz, and broad banded for $\Delta f \geq 0.5$\,Hz. The forcing parameters are summarized in table\ \ref{tabnantes}.

\begin{table}
\centering
\begin{tabular}{c l c c c}
Forcing type & Parameters & Broad & Narrow \\
\hline 
\multirow{3}{5cm}{Filtered random noise (FRN)} & Peak frequency $f_m$ & 1 Hz & 1.15 Hz\\
& Bandwidth $\Delta f$ & 0.6 Hz & 0.3 Hz\\
&  $f_{min}$, $f_{max}$ & 0.7, 1.3  Hz &  1, 1.3 Hz\\
& Wave amplitude $\sigma_{\eta} $ & 0.7 - 6.5 cm & 0.7 - 5.0 cm\\
\hline
\multirow{3}{5cm}{Jonswap (JON)} & Peak frequency $f_m $ & 1 Hz & 1.15 Hz\\
&  Bandwidth $\Delta f$ & 0.6 Hz & 0.4 Hz\\
&  $f_{min}$, $f_{max}$ & 0.7, 1.3 Hz &  0.9, 1.3 Hz\\
& Wave amplitude $\sigma_\eta $ & 1.3 - 3.2 cm & 1.3 - 3.2 cm\\
\hline
\end{tabular}
\caption{Forcing parameters to generate a prescribed spectrum of wavemaker displacement with a spectral shape (JON or FRN), a frequency bandwidth $\Delta f=f_{max}-f_{min}$, and a maximum spectrum amplitude at frequency $f_m$ (see figure \ref{spectresbatteurs}). The corresponding wavelengths are $\lambda_m\approx 1.5$ m (broad) and 1.2 m (narrow), respectively.
}
\label{tabnantes}
\end{table}

The typical power spectrum density of wavemaker displacement is shown in figure~\ref{spectresbatteurs} for a broad band forcing. It is computed from the displacement sensor fixed on top of the wavemaker. No significant change is observed in the spectrum shape when changing the forcing type (JON or FRN), whereas the frequency bandwidth and the amplitude of the spectrum peak are well controlled by $\Delta f$ and $\sigma_{\eta}$, respectively. Thus, the forcing type (JON or FRN) will not be distinguished in the discussion. 

\subsection{Boundary conditions}
Two boundary conditions were tested as illustrated in figure \ref{flumeschema}. First, the wave basin is equipped with an absorbing sloping beach at the opposite end of the basin to the wavemaker, in order to strongly reduce wave reflections. The beach is a porous beach made of stones with a weak slope of the order of 1/3 for the first 3.2 meters, the last 3.5 meters being almost flat. This enables wave absorption by wave breaking and porosity. The amplitude of reflections is estimated to be less than $10\%$ after 5 min of irregular wave generation of peak period of 1 s \citep{Bonnefoy05}. Thus, waves propagate up to the beach with almost no reflections going back ($< 10\%$). This boundary condition will be subsequently referred to as the {\it{absorbing boundary condition}}. The second configuration consists of a wooden wall vertically fixed in the wave basin in front of the beach (see figure~\ref{flumeschema}). This case, called the {\it{reflecting boundary condition}}, corresponds to a closed basin, a situation already tested in previous laboratory experiments on gravity wave turbulence of various basin sizes \citep{Falcon2007,Herbert2010,Cobelli2011,Denissenko2007,NazarenkoJFM2010,NazarenkoAdvance2013,Nazarenkojetp2013}. We will show in the following that the boundary conditions  play an important role on the dynamics of the wave field.

\subsection{Wave gauges}
We use an array of four wave gauges (two capacitives and two resistives) to measure the wave amplitude, $\eta(t)$, as a function of time with a sampling frequency of 500 Hz during typically $T=10$ or 19 min. Resistive gauges are $80$ cm in height. Their vertical resolution is about 0.1 mm, and their frequency resolution is close to $10$ Hz \citep{Bonnefoy05}. The capacitive gauges are $60$ cm in height and are homemade \citep{Falcon2007}. Their vertical resolution is about $0.1$ mm and the frequency resolution up to $200$ Hz. The location of the probe array in the basin is shown in figure \ref{flumeschema}. They are located in the middle of the basin, 7.5 m from the wavemaker, corresponding to a distance of $5\lambda_m$ for the smallest value of $f_m$ used. The measurement can be thus considered ``far" from the wavemaker with respect to previous experiments (see table \ref{tabnantes}) in which the basin size was of the order of $\lambda_m$ \citep{Falcon2007,Herbert2010,Cobelli2011}. Indeed, the forcing scale $\lambda_m$ being not in the inertial range of the cascade (see Sect. \ref{ResSpectre}), one should use a scale $\xi$ at the beginning of the cascade, rather than $\lambda_m$ (e.g. $\xi \simeq17$ cm corresponding to a cascade beginning at 3 Hz - see Sect. \ref{ResSpectre}). The gauges are then located at a distance from the wave maker corresponding to $42$ spatial scales $\xi$ ($L/\xi \simeq 84$). We have also verified that the wave spectrum measured in the vicinity of the wavemaker is different from that measured in far field in the center of the basin. All results obtained here are found to be independent of the gauge type in the working range of the gauges, and of the spectral shapes prescribed to the wavemaker. Moreover, they do not depend significantly on the gauge location within the basin except when the gauges are too close to the boundaries (wavemaker, beach or walls). Typically, $\sigma_{\eta}$ varies less than 5\% for different gauge locations, keeping all the other parameters fixed.

 \begin{figure}
 \begin{center}
\includegraphics[scale=0.4]{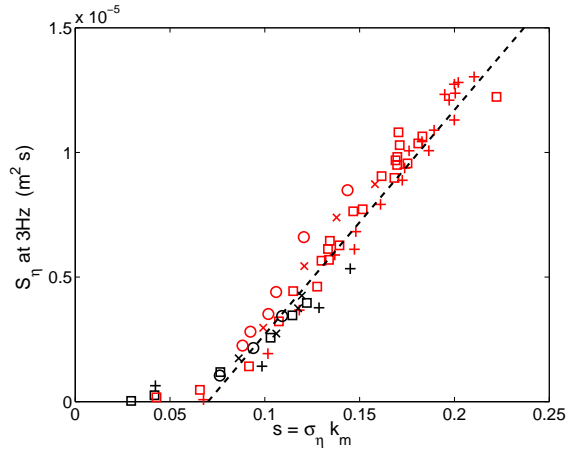} %fig4.eps
 \caption{Amplitude of the wave spectrum amplitude $S_{\eta}(f)$ measured at $f=3$ Hz as a function of mean wave steepness $s$. Dashed line has  slope 1. Forcing parameters: broad bandwidth  ($f_m=1$ Hz, $\Delta f=0.6$ Hz) for FRN ($\square$) or JON ($\circ$) forcings; narrow bandwidth ($f_m=1$ Hz, $\Delta f=0.3$ Hz) for FRN ($+$) or JON ($\times$) forcings. Black: beach. Red (light-grey): Wall.}
    \label{PSD3Hzvsteepness}
       \end{center}
 \end{figure}

\subsection{Wave amplitude parameter}
\label{focingpara}
Several parameters have been used in the literature to quantify irregular wave amplitudes. A natural choice is the rms wave amplitude, $\sigma_{\eta}$, a value directly related to the area under the wave spectrum, $S_{\eta}(f)$. In oceanography, the significant wave height $H_s$ or $H_{1/3}$ was traditionally defined as the mean wave height (trough to crest) of the highest third of the waves. Now, it is usually defined as $H_s=4\sigma_{\eta}$, and with this choice $H_s$ and $\sigma_{\eta}$ are equivalent. The mean injected power by the wavemaker within the system has also been  used previously. However, an unknown amount of energy is injected into the bulk and not into the waves \citep{Deike2014}. The mean wave steepness (or wave slope) $s$ is useful to quantify the degree of nonlinearity of the wave field. It is usually defined as $s\equiv \sigma_{\eta}k_m$, with $k_m$ the wavenumber corresponding to the maximum amplitude of the spectrum. In all our experiments, $k_m$ is roughly constant, and is located in the forcing range $k_{m}=2\upi/\lambda_m\approx 4.2$ m$^{-1}$. The range of the nonlinearity parameter is $0.02 < s < 0.25$. The value of the spectrum amplitude at the beginning of the cascade, but outside the forcing frequency range, is a more relevant parameter \citep{NazarenkoJFM2010}. Since there is no trivial relation between the input energy by the wavemaker, and the energy flux cascading through the wave scales, the spectrum amplitude at the beginning of the cascade is actually a relevant parameter to quantify the magnitude of the cascade of gravity wave turbulence. For instance, for a forcing frequency bandwidth close to 1 Hz, the amplitude of the wave spectrum measured at 3 Hz corresponds roughly to the beginning of the cascade of gravity wave turbulence, and is also well separated from the first harmonic of the forcing. In the following, we will choose the spectrum amplitude at 3 Hz, $S_\eta (3\mathrm{Hz})$, as the parameter that characterizes the forcing amplitude. As shown in figure \ref{PSD3Hzvsteepness}, we found that this parameter increases monotonically with the mean wave steepness but not with a simple scaling (nonlinearly at small $s$, then linearly at high $s$). However, this relationship does not depend on the basin boundary conditions.

\section{Role of basin boundary conditions on the wave field}
\label{reel}
We focus here on the influence of the basin boundary conditions on the wave field in real space.

\begin{figure}
\centerline{
\begin{tabular}{c}
\includegraphics[scale=0.4]{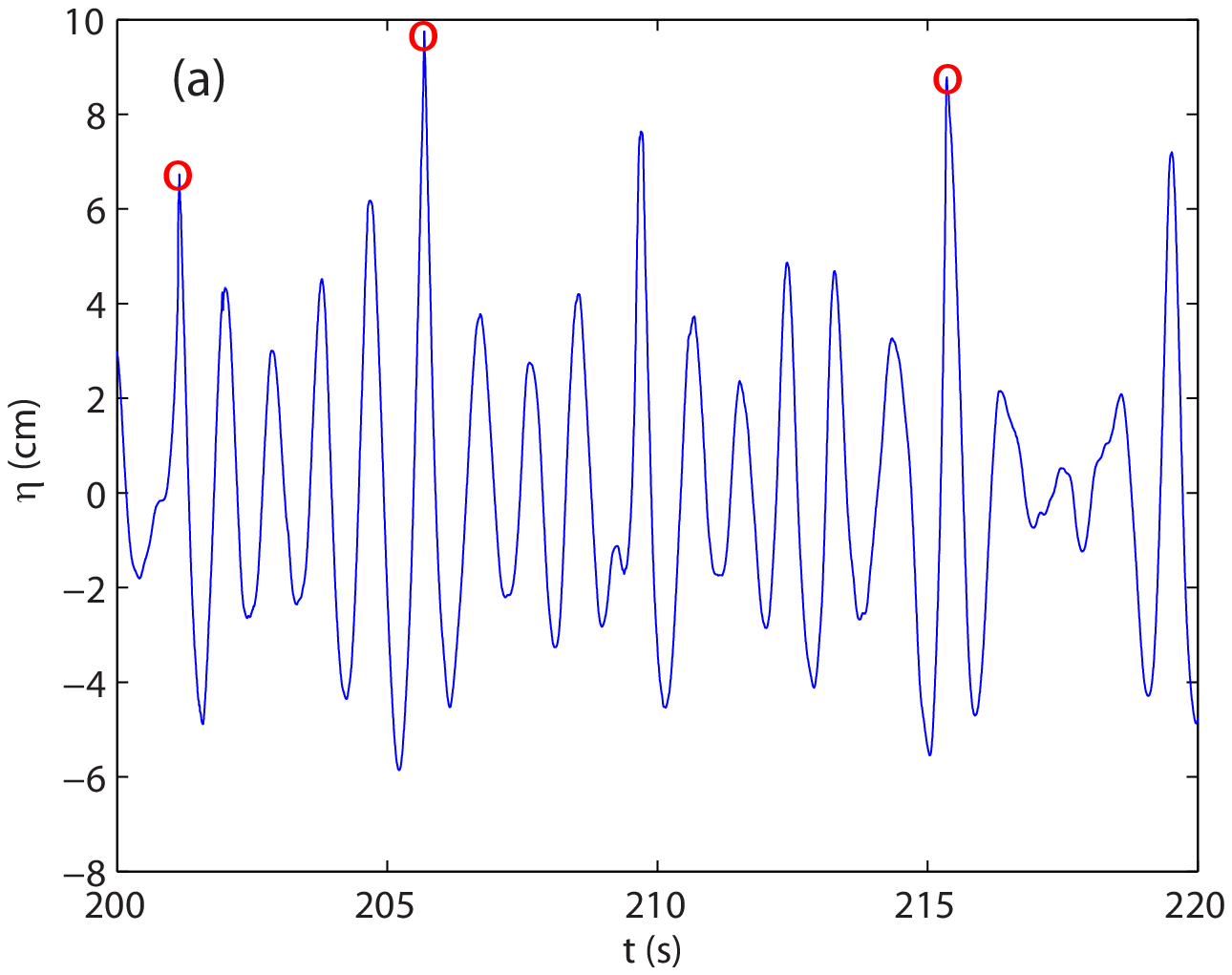} %fig4beach.eps
\includegraphics[scale=0.25]{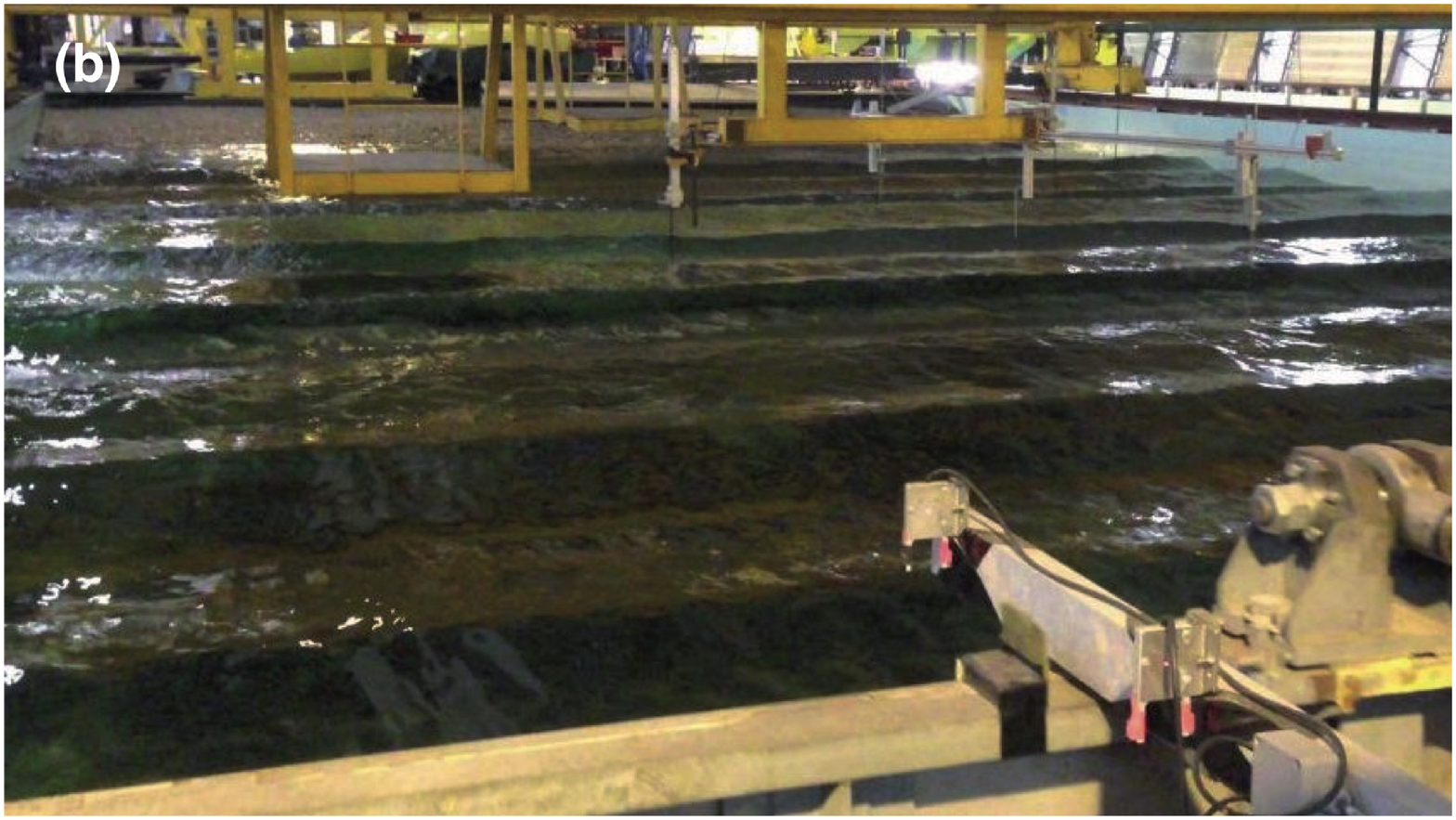}\\ \\ %PlageBB4IMG39011Avril2014bis
\includegraphics[scale=0.4]{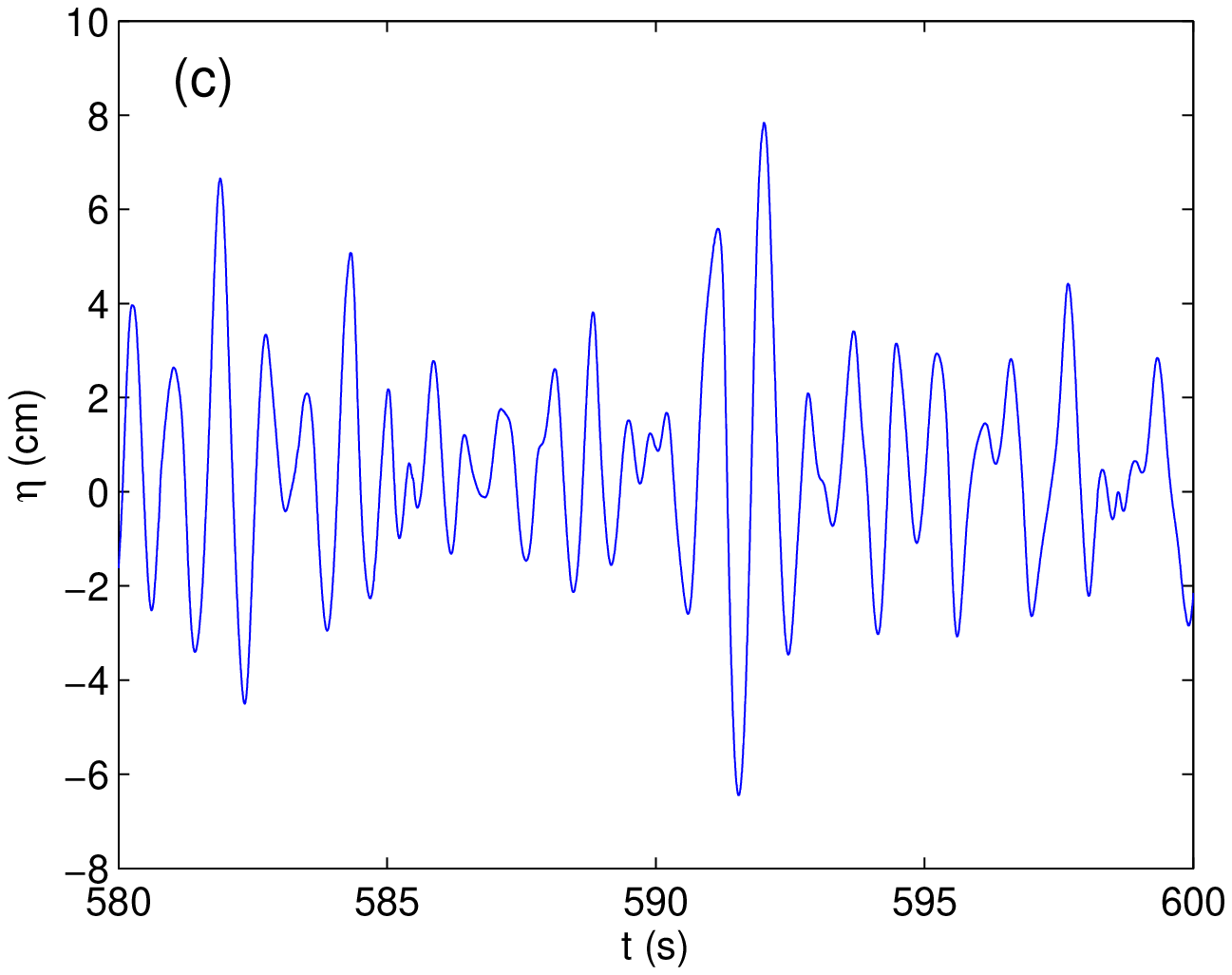} %fig4wall.eps
\includegraphics[scale=0.25]{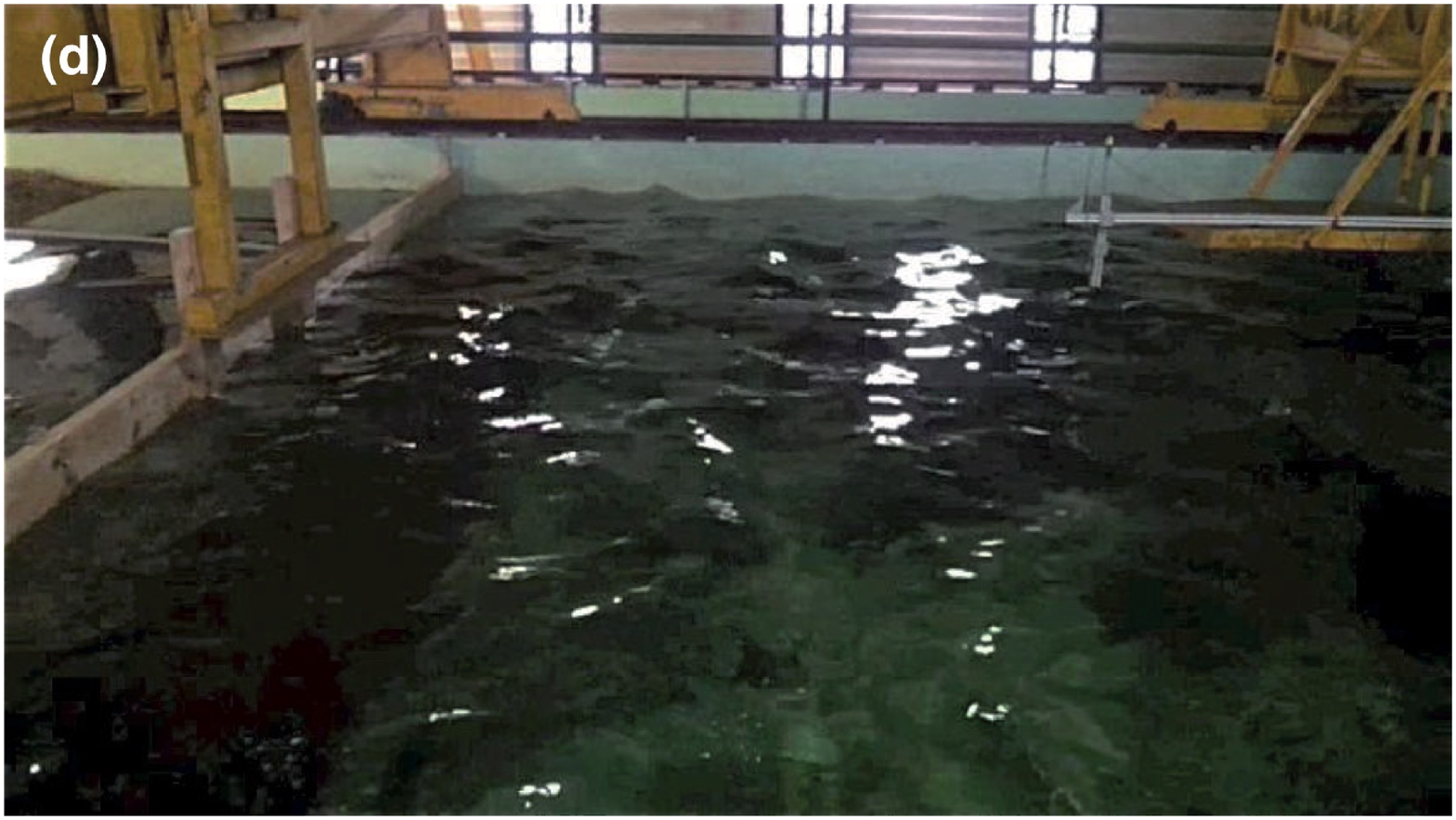} %ImageMurBB20new2.eps
\end{tabular}
}
\caption{Typical temporal evolution of the wave amplitude $\eta(t)$ (left), and the corresponding wave field picture (right) for two different boundary conditions: beach [top - (a) and (b)] or wall [bottom - (c) and (d)]. The array of probes, the beach, and a part of the wavemaker are visible in (b). The wall is visible in (d), the shot angle being different. $\circ$-marks in (a) corresponds to sharp crest events. Broad bandwidth ($f_m=1$ Hz, $\Delta f=0.6$ Hz) for FRN forcing. $\sigma_{\eta}=2$ cm (top) and 2.9 cm (bottom).}
\label{Photo}
\end{figure}

\subsection{Direct observation of the wave field}
Irregular waves (of random frequency and amplitude) are generated by the wavemaker as explained in Sect. \ref{setup}. The typical temporal evolution of wave amplitude $\eta(t)$ and the corresponding picture of the wave field in a stationary regime are shown in figure \ref{Photo} for two different boundary conditions: beach (top) or wall (bottom). Direct observation of the wave field shows that its spatial structure depends strongly on the absorbing or reflecting boundary condition. In the absorbing case (beach), a quasi-one dimensional field of nonlinear waves propagates from the wavemaker before being damped by the beach (see figure \ref{Photo}b). In the reflecting case (wall), such coherent structures are not visible. Instead a multidirectional wave field is observed (see figure \ref{Photo}d) due to nonlinear interactions between waves and multiple refleions occurring from the basin walls. Note that the direction of forcing is one-dimensional in both cases, and the wave steepnesses are of the same order. The temporal evolution of the wave amplitude $\eta(t)$ is shown in figures \ref{Photo}a and \ref{Photo}c. Both signals are erratic showing rare large wave events, as well as higher frequency components than the forcing ones. Note that sharp crest events seem more probable in the beach case (as emphasized by circles in figure \ref{Photo}a), occurring only rarely with a wall (figure \ref{Photo}c). The displayed sample for each boundary condition is representative of the whole time series. Moreover, for the highest forcing amplitudes, we occasionely observe the presence of breaking events during the propagation similar to those studied in laboratory flumes~\citep{MelvilleJFM82,Perlin2013}. At sufficiently high  forcing amplitude and for both boundary conditions, we find that the probability distribution function of wave amplitude is well described by a Tayfun distribution (the first quadratic nonlinear correction to the Gaussian) \citep{Tayfun1980,Socquet2005} as already observed in laboratory experiments \citep{Onorato04,Falcon2007,Onorato09,Falcon2011} or in oceanography \citep{Ochi1998,Forristall2000}. 
 
\subsection{Spatial correlation of the wave field}
\label{corr}
We compute the spatial correlation between the wave gauges to quantify the basic spatial properties of the wave field. The correlation between the wave gauge $i$ and $j$ reads $C_{ij} (\tau) =  \lim_{T\rightarrow \infty} \frac{1}{T}\int_{-T}^{T} \eta_i (t) \, \eta_j (t+\tau) \, dt/\sqrt{C_{ii}\,C_{jj}}$ where $C_{ii}=\lim_{T\rightarrow \infty} \frac{1}{T}\int_{-T}^{T} \eta_i (t) \eta_i (t+\tau) \, dt $ is the autocorrelation function. The correlation function is thus normalized between $-1$ and $1$. The maximum over the time $\tau$ of the correlation, $C^m_{ij}$, gives information on the wave field mixing and propagation properties between the gauges. $C^m_{ij}=1$ occurs when signals from gauges $i$ and $j$ are totally correlated, while $C^m_{ij}=0$ corresponds to two signals completely uncorrelated. Note that the correlation between two gauges depends on the linear dispersion of a wave packet, the propagation direction of the waves, as well as decorrelation induced by nonlinear interactions.

\begin{figure}
 \centering
\includegraphics[scale=0.4]{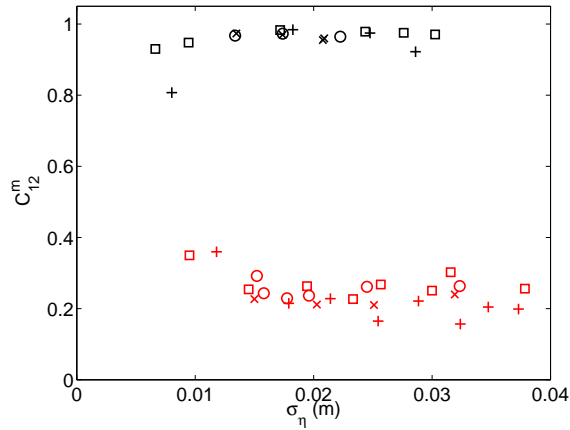} %fig3b.eps
  \caption{Maximum of correlation amplitude, $C_{1,2}^m$, between the wave gauges 1 and 2 as a function of rms wave amplitude, $\sigma_\eta$, for two different boundary conditions:  absorbing (beach - black) and reflecting (wall - red light grey). Symbols correspond to the same forcing parameters as in figure \ref{PSD3Hzvsteepness}.}
   \label{Correlation}
 \end{figure}  

Figure \ref{Correlation} shows the maximum of correlation amplitude, $C_{1,2}^m$, between two probes facing the wavemaker (wave gauges C1 and C2). These probes are located at the same distance from the wavemaker and are separated by 2 m (see figure \ref{flumeschema}). The maximum correlation is reached for $\tau\simeq 0$. We found that $C_{1,2}^m$ depends strongly on the basin boundary conditions. For the absorbing condition (beach), the wave amplitudes are highly correlated whatever the forcing ($C_{1,2}^m$ close to 1), while for the reflecting boundary condition (wall), the correlation is low ($C_{1,2}^m < 0.4$). A two-point correlation close to 1 means that the same wave train is observed at the two probes at the same time. This confirms quantitatively the fact that, in the case of the beach, the wave field remains almost one-dimensional during the propagation. For the wall case, the correlation is much lower due to the multiple reflexions occurring on the basin walls enhancing nonlinear wave interactions. The resulting wave field is thus more complex than observed in figure \ref{Photo}d. The two-point correlation thus confirms direct observation of the wave field pictures. Note that similar results are found for the correlation between two probes aligned with the forcing direction. These spatial properties obtained from temporal measurements (even if spatio-temporal ones should be ideally obtained) are mainly related to the forcing properties and to the boundary conditions. We have to keep in mind these simple spatial properties when discussing the wave spectrum in Sect.\ \ref{SSpectre}.  

\section{Role of basin boundary conditions on the wave spectrum}
\label{SSpectre}
We now discuss the role of the boundary conditions on the wave field in the Fourier space.

\subsection{Wave spectrum}\label{ResSpectre}
\begin{figure}
 \begin{center}
\includegraphics[scale=0.4]{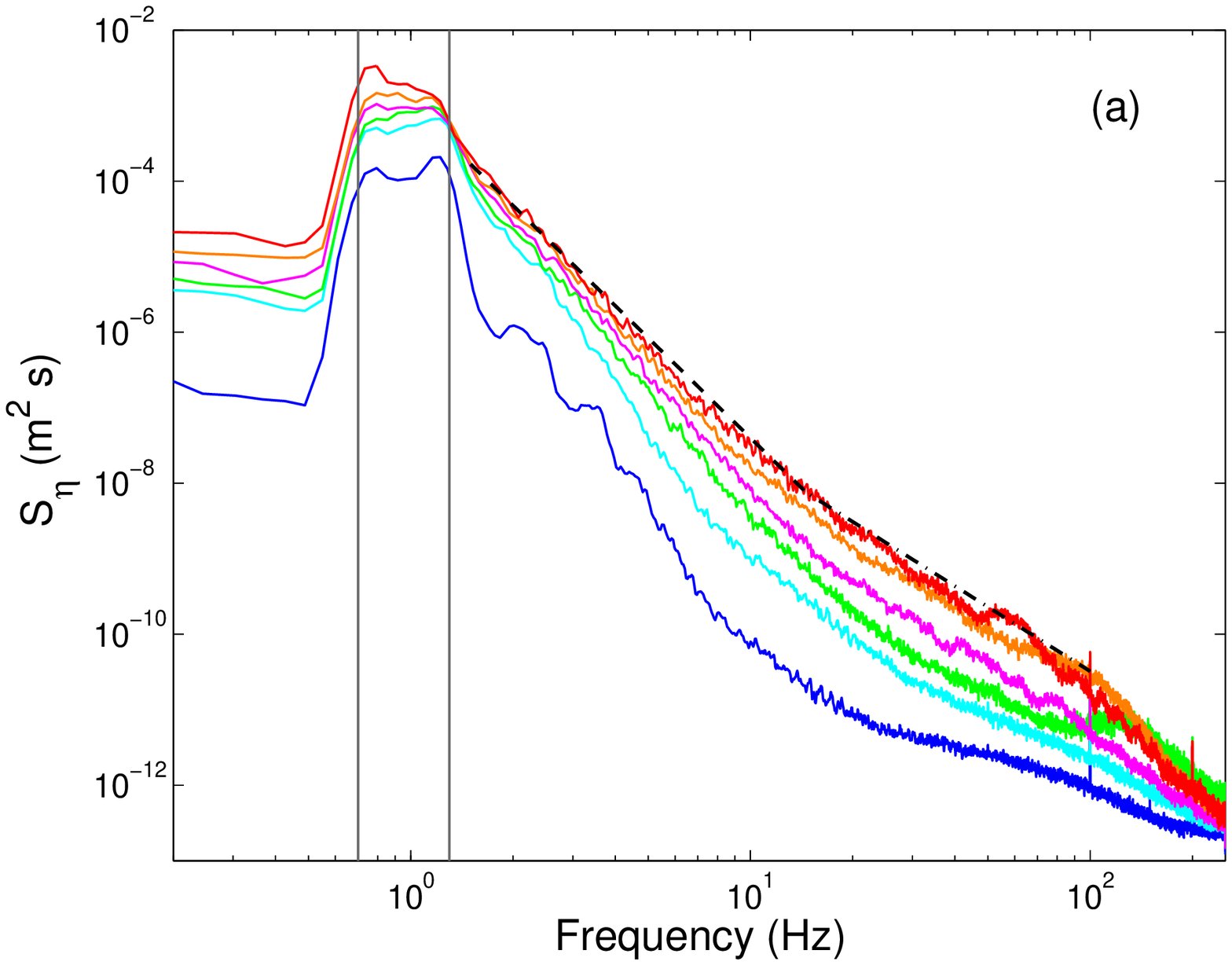} %fig5.eps
 \includegraphics[scale=0.4]{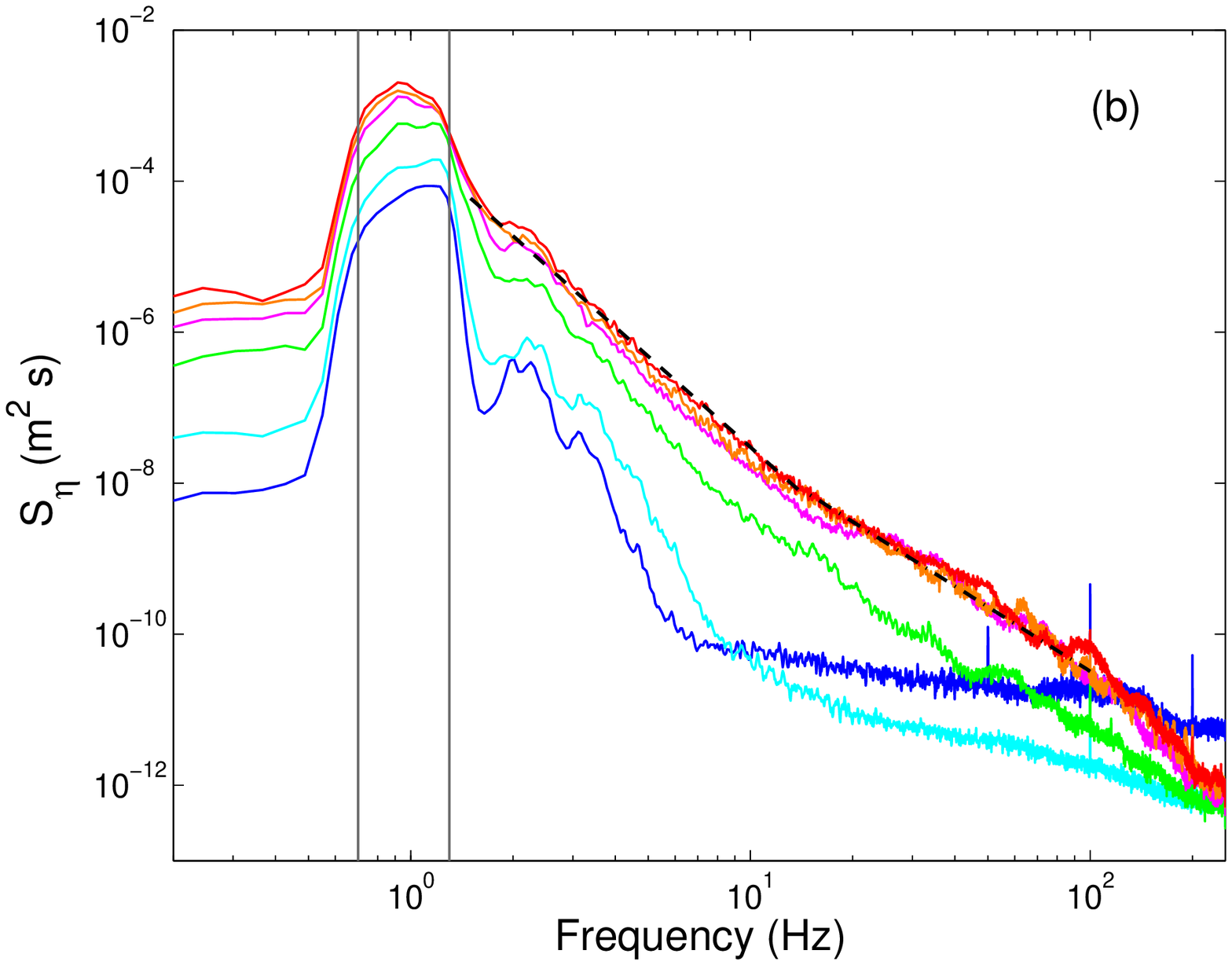} %fig6.eps
 \caption{Power spectra of wave amplitude, $S_{\eta}(f)$, for reflecting (a) or absorbing (b) boundary conditions. Forcing amplitude increases from bottom to top. (a) Weak turbulence predictions for gravity regime $S_{\eta} \sim f^{-4}$ (dashed line) and capillary regime $S_{\eta} \sim f^{-17/6}$ (dot-dashed line). (b): Best fit $\sim f^{-4.4}$ in the gravity regime (dashed line), and $f^{-17/6}$ in the capillary regime (dot-dashed line). Vertical grey lines indicate the forcing frequency range. The same forcing parameters were used for (a) and (b). FRN forcing with broad bandwidth ($f_m=1$ Hz, $\Delta f=0.6$ Hz, and $0.6 \leq \sigma_{\eta} \leq 3.7$ cm).}
    \label{compspWN}
       \end{center}
 \end{figure}

Figure \ref{compspWN} shows the wave amplitude spectra, $S_{\eta}(f)$, for increasing forcing amplitudes for reflecting (a) or absorbing (b) boundary conditions. Surprisingly, both conditions lead to the same qualitative shape of the spectra as the forcing is increased. For small forcing amplitude, peaks related to the forcing and its harmonics are visible in the low frequency part of the spectrum and no power law is observed. At sufficiently high  forcing, those peaks are smoothed out and a power law, $S_{\eta}(f)\sim f^{\alpha}$, can be fitted. This corresponds to the cascade of gravity wave turbulence over a one decade  frequency-range from roughly $1.5$ Hz (the higher forcing frequency) up to the gravity-capillary crossover frequency $f_{gc}\equiv \sqrt{2g/l_c}/(2\upi)\simeq 14$ Hz with $l_c\equiv \sqrt{\gamma/(\rho g)}$ the capillary length, $g=9.81$ m/s$^2$ the acceleration due to gravity, $\gamma=70$ mN/m the surface tension, and $\rho=1000$ kg/m$^3$ the water density \citep{Falcon2007}. When the forcing is further increased, the slope of the power law spectrum becomes less steep, corresponding to an increase in the exponent $\alpha$. Finally, for the highest forcings, the slope seems to saturate to a constant value (see dashed line) although the peak amplitude of the forcing frequencies still increases. For both boundary conditions, this value is close to $-4$ the exponent predicted by gravity wave turbulence theory \citep{Zakharov1967}. This may be coincidental, since effects of dissipation and nonlinear coherent structures are strongly involved experimentally but are not taken into account in weak turbulence theory (see Sect. \ref{diss}). Note that the role of dissipation has recently been studied theoretically \citep{ZakharovPRL07}.

Let us now look at the high frequency part of the spectrum, corresponding to the capillary range, where the spectrum shape changes. At sufficiently high forcing, a second power law is indeed observed over a one-decade frequency-range ($f_{gc} < f < 100$ Hz). The slope is much less steep than that for the gravity range and is close to the capillary wave turbulence prediction in $f^{-17/6}$ (see dot-dashed lines) \citep{Zakharov1967cap}. Note that the observation of both direct cascades of gravity and capillary wave turbulence was practically unattainable in previous large basin facilities. It is possible here due to both the high sensitivity and low noise level of the capacitive probes, the latter being reached for $f \gtrsim 200$ Hz. These results are found to be independent of the forcing parameters (spectral shape and frequency bandwidth).

As discussed in Sect. \ref{reel}, the propagation of a quasi one-dimensional field of nonlinear waves is observed in the presence of a beach, whereas the presence of a wall leads to numerous propagation directions and consequently a multidirectional wave field. Although the different boundary conditions yield pronounced differences in wave field structure there is surprisingly no significant difference in the corresponding wave spectra.

\begin{figure}
 \begin{center}
\includegraphics[scale=0.4]{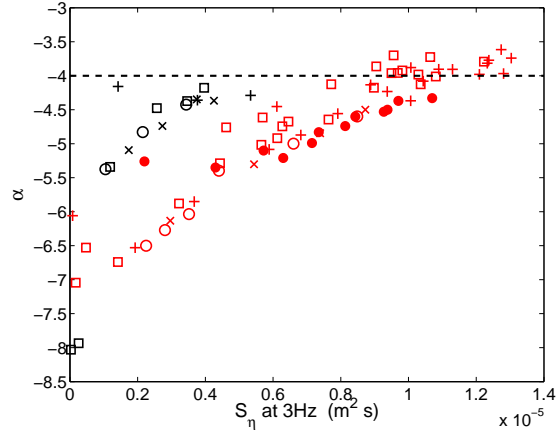}  %fig7.eps
\caption{Frequency power-law exponent, $\alpha$, of the gravity wave spectrum $S_{\eta}\sim f^{\alpha}$ as a function of the value of the spectrum amplitude at 3 Hz, $S_{\eta}(3{\ \rm Hz})$. Boundary conditions: wall (red light grey) and beach (black). Dashed line corresponds to the prediction of weak turbulence theory $\alpha=-4$. Fit frequency range: 3 to 10 Hz. Error bar on $\alpha$ is $\pm 0.2$. Symbols correspond to the same forcing parameters as in figure \ref{PSD3Hzvsteepness} and $0.5 < \sigma_{\eta} < 5$~cm. ($\bullet$)-symbols correspond to Hull experiments of figure 6 (left) in \cite{NazarenkoJFM2010}.} 
    \label{exposant}
       \end{center}
 \end{figure}

The frequency power law of the gravity wave spectrum, $S_{\eta}(f)\sim f^{\alpha}$, is found to depend on the forcing amplitude. Figure \ref{exposant} shows $\alpha$ as a function of the forcing strength for both the absorbing and reflecting boundary conditions. We choose to plot it as a function of $S_{\eta}(3{\ \rm Hz})$, the value of the spectrum amplitude at 3 Hz (a forcing strength parameter more relevant than the mean wave steepness $s$, or the rms wave amplitude $\sigma_{\eta}$ as explained in Sect. \ref{focingpara}). The exponent $\alpha$ is found to increase with the forcing strength for both boundary conditions. In a closed basin, $\alpha$ seems to saturate at high forcing near $-4$ within the data scattering. In the presence of a beach, the highest value reached by $\alpha$ is also $-4$ but occurs at a smaller $S_{\eta}(3{\ \rm Hz})$. For both boundary conditions, $\alpha$ is thus found to be close to $-4$ at sufficiently high forcing. The maximum value of $S_{\eta}(3{\ \rm Hz})$ reached in the presence of a beach is less than that obtained with a wall for the same forcing parameters. This arises from the fact that the dissipated power is stronger in presence of a beach than within a closed basin. As stated earlier, our results are  independent of the spectral shapes prescribed to the wavemaker. Finally, when comparing our results performed in the closed basin with those reported in the Hull experiments \citep{NazarenkoJFM2010} [see ($\bullet$)-symbols in  figure \ref{exposant}], a good overall agreement is found although a smaller value of $S_\eta(3{\ \rm Hz})$ is needed in our case to reach the same value of $\alpha$.

\subsection{Discussion}
\label{diss}
The weak turbulence prediction for the wave spectrum in the gravity regime reads $S_{\eta}(f) \sim f^{-4}$ \citep{Zakharov1967}, and is depicted by a dashed line in figure \ref{exposant}. It seems to roughly describe the data at sufficiently high  $S_{\eta}(3{\ \rm Hz})$ for both boundary conditions. However, one would have expected a better agreement with the data at low $S_{\eta}(3{\ \rm Hz})$, i.e. at low wave steepness, since this theory is weakly nonlinear. Moreover, a lower $S_{\eta}(3{\ \rm Hz})$ is needed to reach this $-4$ value in the presence of a beach than of a wall. This is somewhat paradoxical since the spatial structure of the wave field in the presence of a beach involves mainly unidirectional coherent structures (see figure \ref{Photo}b) whereas a multidirectional wave field is observed with a wall (see figure \ref{Photo}d), this latter situation being much closer to the isotropic assumption of weak turbulence. However, some very small correction to the $-4$ exponent is predicted due to anisotropy \citep{Pushkarev03}, and generally swell and blowing wind are barely isotropic in oceans. Moreover, weak turbulence also assumes uncorrelated waves between two distant points. We observed that in the presence of the wall, signals of different probes are uncorrelated at the scales within the turbulent cascade. In contrast, with the beach, a significant correlation remains, suggesting again that coherent structures play an important role. Note that weak turbulence also predicts the existence of an inverse action cascade of gravity waves with a power spectrum in $f^{-11/3}$, close to the direct energy cascade exponent $-4$. The two regimes are thus hardly distinguishable experimentally here. However, no inverse cascade of gravity waves is observed from our large scale forcing towards larger scales. Indeed, our forcing scale (near 1 Hz) is too close to the largest achievable scale (0.35 Hz) corresponding to a wavelength equal to the basin size. Coexistence of both cascades within the inertial frequency range could be yet possible but is unlikely for weak wave steepnesses where no nonlocal forcing at small scales is expected. Note that inverse cascade of gravity waves has been recently observed by injecting energy at an intermediate scale corresponding to the gravity-capillary length \citep{Deike2011}.

A possible explanation of the $f^{-4}$ spectrum scaling at sufficiently high forcing is given by te spectrum of one-dimensional spatial singularities \citep{Kuznetsov2004,NazarenkoJFM2010}. If the wave field dynamics is dominated by 1D sharp crested waves propagating with a preserved shape, as observed in the beach case, the Fourier transform of the amplitude of these singularities is $\hat{\eta}(k) \sim k^{-2}$. Its power spectrum is $S_{\eta}(k)\sim |\hat{\eta}(k)|^2 \sim k^{-4}$ in wavenumber, and $S_{\eta}(\omega) = S_{\eta}(k) \frac{dk}{d\omega} \sim \omega^{-4}$ in frequency, assuming a constant group velocity (i.e. $\omega \sim k$). However, our temporal measurements of the wave amplitude with a probe at a single location cannot discriminate which mechanism is involved at high forcing, either the singular coherent structures or the resonant wave interactions of weak turbulence theory. For this, full space and time resolved measurements of wave elevations are needed since coherent structures do not belong to the linear dispersion relation curve and thus should be easily detectable. A spatio-temporal measurement of wave height working in the gravity range could be tested, similar to measurements used for gravity-capillary wave turbulence \citep{Herbert2010,Cobelli2011}, capillary wave turbulence \citep{Putterman1996,Berhanu2013}, or hydrodynamics surface waves \citep{Zhang2002,Cobelli2009}. Note that within our experimental setup, it is not possible to perform spatio-temporal measurements as in \cite{Herbert2010,Cobelli2011}. This is because white liquid dye cannot be added to water to improve its light diffusivity due to basin guidelines. Other methods measuring the surface gradient of the wave field both in space and time \citep[see][and references therein]{Moisy2009} are intrinsically limited to weak wave steepness and hence are of limited usefulness here.

One way to interpret our results at high forcing would be to ascribe the observed spectra to the propagation of coherent structures in the presence of a beach and to a weak turbulence mechanism in the presence of a wall. However, this does not explain the spectrum exponent dependence on the forcing in both cases (see figure \ref{exposant}). It has been shown previously that  removing such coherent structures from the wave amplitude signal leads to a gravity spectrum exponent that still depends on the forcing but with less variation, of the order of 25\% \citep{Falcon10b}. Using a similar criterion to define the occurrence of wavebreaking events (time intervals where the wave acceleration is greater than six times its standard deviation),  compute the spectrum of the wave signal not including wave breakings. We found that the spectrum exponent is only decreased by roughly 10\% but still depends on the forcing, and no clear difference is observed between the wall and beach cases within our data scattering.

The relative importance of dissipation (e.g. by wave breaking) with respect to nonlinear interactions may also explain the steepening of the gravity spectrum at low nonlinearity. In capillary wave turbulence, a similar phenomenon of steepening of the spectrum at low nonlinearity has been reported experimentally when working with fluids of sufficiently high viscosity  \citep{Deike2014} and numerically when reducing the nonlinear interactions \citep{Pan2014}. 

To summarize, we have observed gravity-wave turbulence spectra that present strong discrepancies with weak turbulence theory. There are possible physical effects responsible for these differences, that are usually not taken into account theoretically: presence of nonlinear coherent structures, anisotropy of the wave field, and dissipation at all scales of the cascade. More specifically, we observe for some experiments a frequency spectrum exponent equal to $-4$, which is the exponent predicted by weak turbulence theory for gravity waves. However, this exponent depends on the forcing amplitude, and the value $-4$ is reached at lower forcing in the presence of a beach (involving quasi one-directional waves) than in the presence of a wall (where the wave field is multidirectional). This discrepanc is probably due partly to the propagation of nonlinear coherent structures, and mainly to wideband dissipation. Finally, note that widening of the wave dispersion relation due to nonlinearities has been shown to permit one-dimensional wave interactions \citep{AubourgPRL2015}. To what extent, a similar 1D mechanism is relevant in the beach case remains an open question that warrants further study.  

The next section dealing with intermittency in wave turbulence may give insights into the mechanisms in play.

\section{Role of basin boundary conditions on intermittency}\label{inter}
The phenomenon of intermittency has been observed experimentally in gravity wave turbulence \citep{Falcon07b,Falcon10b, NazarenkoJFM2010}. Here, we investigate the role of the boundary conditions on the intermittency properties in gravity wave turbulence. 

The intermittence of a stochastic stationary signal, $\eta(t)$, is generally tested by computing the structure functions using the first-order differences of the signal, $\eta(t+\tau)-\eta(t)$. A signal with a steep power spectrum, $S_{\eta}(f) \sim f^{\alpha}$, is locally multi-derivable, and  high-order difference statistics is then required to test intermittency \citep{Falcon10a,Falcon10b}. For instance, with $| \alpha | \geq 5$, at least third-order difference statistics is required. Here, we find that statistical convergence of the structure functions is reached when using the fourth-order (or higher) difference statistics. The fourth-order differences of the signal
 ${\Delta}\eta_t(\tau)\equiv {\eta}(t+2\tau)-4{\eta}(t+\tau)+6{\eta}(t)-4{\eta}(t-\tau)+{\eta}(t-2\tau)$, are thus computed in the following. $\eta(t)$ is recorded with a 500 Hz sampling rate during 19 min leading to $6 \times 10^{5}$ points.
 
\begin{figure}
 \centerline{
\begin{tabular}{cc}
\includegraphics[scale=0.37]{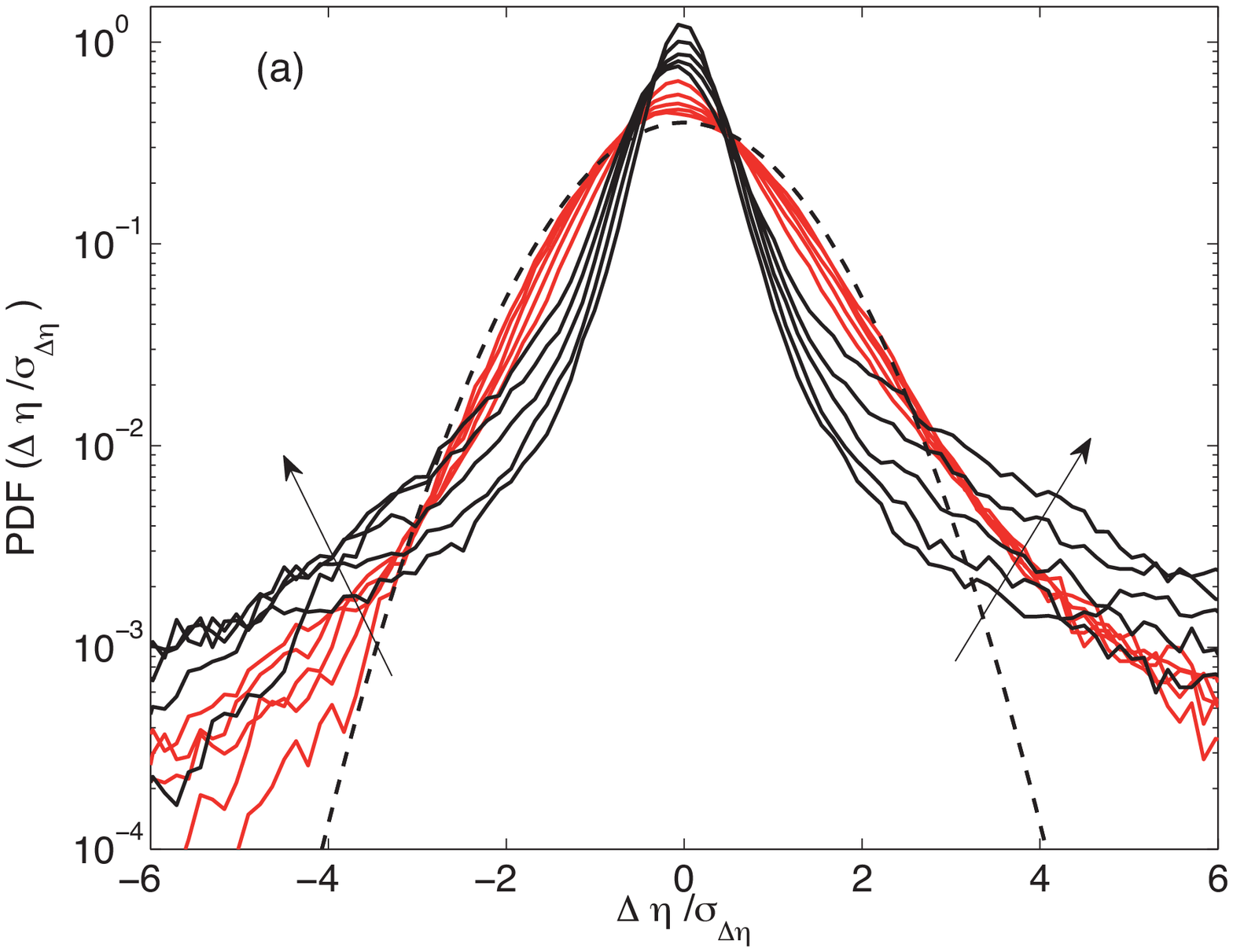} 
\includegraphics[scale=0.37]{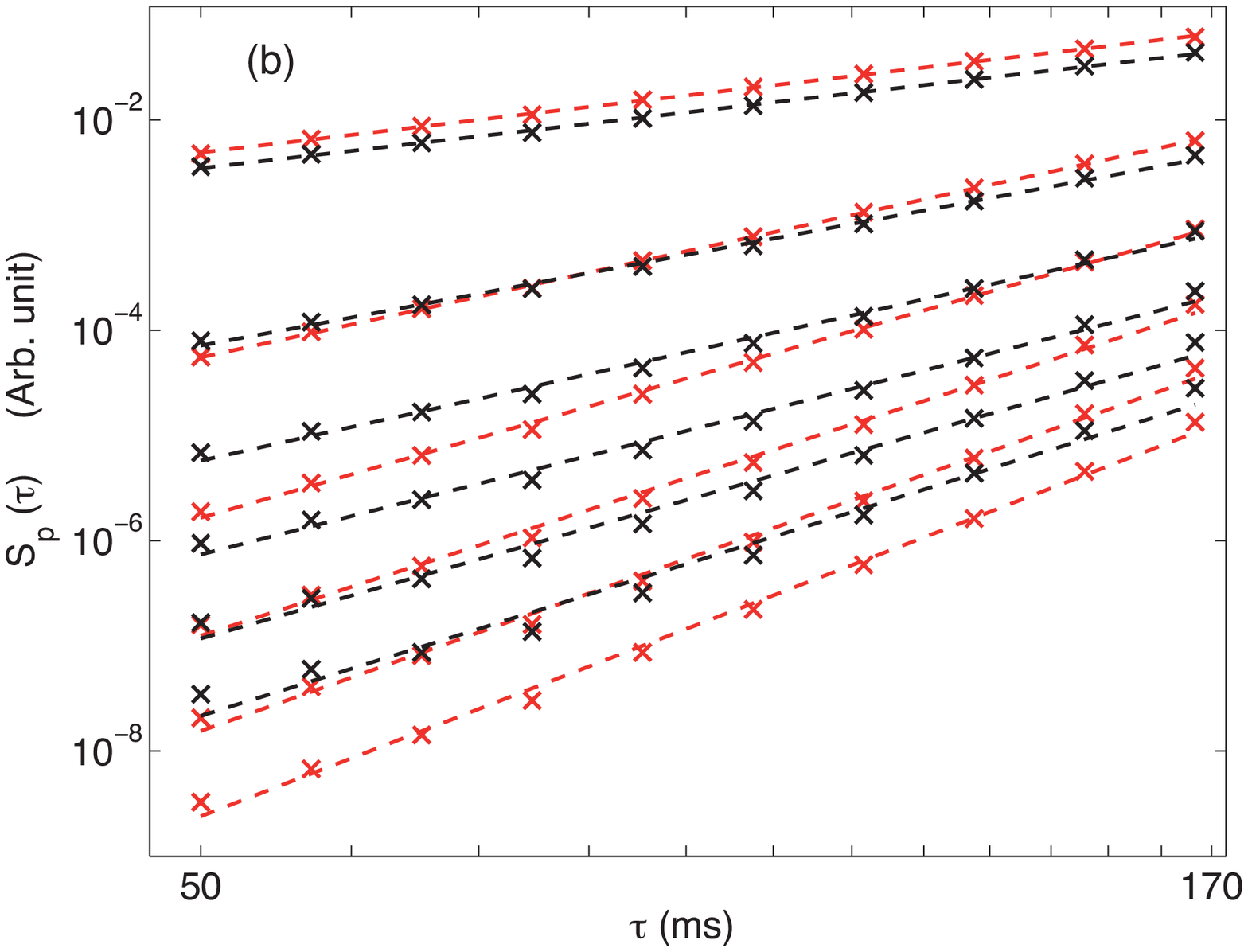} 
\end{tabular}
} %\includegraphics[scale=0.4]{fig08.eps} %figinter1.eps
 \caption{(a) Probability density functions of normalized increments $\Delta \eta_t(\tau)/\sigma_{\Delta \eta_t(\tau)}$ for different time lags $\tau=50$, 65, 85, 111, and 146 ms (see arrows) and two configurations: wall (red light grey) and beach (black). Dashed line: Gaussian with zero mean and unit standard deviation. (b) Structure functions of the fourth-order differences of the wave amplitude, ${\mathcal S}_p \sim \tau^{\zeta_p}$, as functions of the time lag $\tau$, for $1\leq p \leq 6$ (from top to bottom). Wall (red light grey) and beach (black). Dashed lines are corresponding power law fits in which their slopes $\zeta_p$ depends on the order $p$ (see figure \ref{fig2inter}). The same forcing parameters were used as in figures \ref{Photo} and \ref{compspWN} with $\sigma_{\eta} =$3 cm [$S_{\eta}(3 \rm{\ Hz})= 10^{-5}$ m$^2$s (wall) and 0.4 10$^{-5}$ m$^2$s (beach)].}
    \label{fig1inter}
 \end{figure}

The probability density functions (PDFs) of ${\Delta}\eta_t(\tau)$ normalized to their rms values $\sigma_{{\Delta}\eta}$ are displayed in figure \ref{fig1inter}a for different time lags $\tau$ and for two configurations (wall and beach). We choose the range $50 \leq \tau \leq 170$ ms corresponding to a frequency range [$2.9 \leq 1/(2\tau) \leq 10$ Hz] within the gravity regime where the wave spectrum is found to scale as a frequency power law, $S_{\eta}(f) \sim f^{\alpha}$, with $\alpha=-5$ and $-4.2$ for the wall and the beach, respectively. In both cases, we observe that the PDF shape changes continuously when $\tau$ is decreased (see arrows), with smaller scale $\tau$ yielding a more flattened PDF. More intense and rare events  occur in the signal at such shorter time scales, meaning that the PDFs are more intermittent. Two other observations can be made. First, in both cases, the PDFs are not Gaussian at large $\tau$ meaning that intermittency already takes place at the forcing scales. Secondly, in both cases, the PDFs are asymmetric with more positive events than negative ones. This could be ascribed to  wave asymmetry (the shape of the leading wavefront is different to the rear wavefront) due to nonlinear effects. Finally, it can be observed that the tails of PDFs are more populated in the presence of a beach than with a wall, for the same rms wave amplitude, whereas its center is much more peaked. 

\begin{figure}
 \begin{center}
 \includegraphics[scale=0.4]{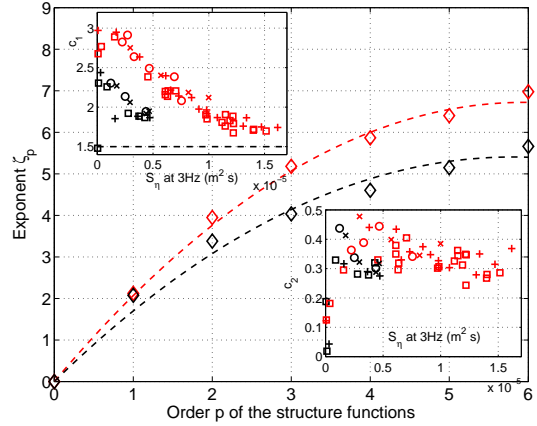} %figinter2.eps
 \caption{Exponents $\zeta_p$ of the structure functions as a function of $p$ for wall (red light grey) and beach (black) configurations. Dashed lines are best fits $\zeta_p=c_1p-\frac{c_2}{2}p^2$. $\zeta_p$ are inferred from the slopes of the power-law fits in figure \ref{fig1inter}b. Top and bottom insets show the evolution of $c_1$ and $c_2$ with the forcing. Symbols in the insets correspond to the same forcing parameters as in figure \ref{PSD3Hzvsteepness} and $0.5 < \sigma_{\eta} < 5$~cm [$S_{\eta}(3 \rm{\ Hz})<$ 1.6 10$^{-5}$m$^2$s].}
        \label{fig2inter}
       \end{center}
 \end{figure}

To quantify the intermittency, the structure functions of order $p$, ${\mathcal S}_p(\tau) \equiv  \overline{ |\Delta\eta_t(\tau)|^p }$, are computed from the fourth-order differences of the signal. ${\mathcal S}_p(\tau)$ are shown in figure \ref{fig1inter}b for both the wall and beach cases, and for comparable $S_{\eta}(3 \rm{\ Hz})$. All the structure functions of order $p$ (from 1 to 6) are well fitted by power laws of $\tau$, ${\mathcal S}_p(\tau) \sim \tau^{\zeta_p}$, where $\zeta_p$ is found to increase with the order $p$ in both cases. The exponents $\zeta_p$ of the structure functions are then plotted in figure \ref{fig2inter} as a function of $p$. $\zeta_p$ is fitted by a quadratic function of $p$ such that $\zeta_p=c_1p-\frac{c_2}{2}p^2$ where the values of $c_1$ and $c_2$ are found to both depend on the forcing (see top and bottom insets of figure \ref{fig2inter}). The $c_1$ coefficient is found to decrease from 3 to 1.7 for increasing $S_{\eta}(3 \rm{\ Hz})$, and to depend on the boundary conditions (see top inset of figure \ref{fig2inter}). This decrease of $c_1$ is due to the decrease of the wave spectrum exponent $| \alpha |$ with $S_{\eta}(3 \rm{\ Hz})$ (see figure \ref{exposant}), since both values are related by $| \alpha | =\zeta_2+1=2(c_1-c_2)+1$. This argument also explains the deviation between the evolutions of $c_1$  in the case of a wall or a beach (top inset of figure \ref{fig2inter}). The nonlinearity of $\zeta_p$ ($c_2\neq 0$) is a signature of intermittency \citep{Pope}. The so-called intermittency coefficient $c_2$ is found to increase from 0 to roughly 0.4 when the forcing is increased. However, no significant difference is observed, within our data scattering, in the presence of a wall or a beach. Similar results have been found when the forcing is increased.

\cite{NazarenkoJFM2010} suggested that instead of fitting $\zeta_p$ by a quadratic function, $\zeta_p$ can be adjusted, for high value of $p$, with a linear fit to measure the fractal dimension of possible singularities involved in the wave field. For the data in the main figure \ref{fig2inter}, a linear fit of $\zeta_p$ for $p>2$ leads to a slope of 0.55, and y-intercept less than 2 in both cases (wall and beach). The fractal dimension inferred from these values and using Eq. (2.28) of \cite{NazarenkoJFM2010} is negative, and thus raises doubts about the validity of this approach.

To conclude, we have found that the intermittency coefficient has roughly the same value in the presence of a beach or a wall, but is found to depend strongly  on the forcing as previously reported \citep{Falcon10b}.  Since it has been shown that intermittency is enhanced by coherent structures \citep{Falcon10b}, our observations  suggest that the importance of coherent structures increases with the forcing both for the beach and the wall with the same trend. The main difference relates to the PDF of increments which displays more rare and intense events in the presence of a beach, and a much more pronounced central peak than with a wall. This probably suggests that the mixing of waves is less efficient, and intense coherent structures are more probable in the presence of a beach than with a wall.
 
\section{Decaying gravity wave turbulence in the closed basin}
\label{decay}
We present here an investigation of freely decaying gravity wave turbulence in the closed basin. Previous experimental studies of such non-stationary regimes have shown that the wave spectrum decays first rapidly as a time-power law in rough agreement with weak turbulence theory, and then exponentially over a longer time interval due to linear viscous dissipation \citep{NazarenkoAdvance2013,Nazarenkojetp2013}. Direct numerical simulations of the Euler equations have also been performed in the freely decaying case of a swell wave field to show the validity of weak turbulence derivation \citep{Onorato02}.

\subsection{Experimental protocol}
We use the same protocol as in previous studies on freely decaying wave turbulence on thin elastic plates \citep{Miquel2011,Deike2013} or on the surface of a fluid \citep{Deike2012,Nazarenkojetp2013,NazarenkoAdvance2013}. A typical experiment is as follows. First, surface waves are generated during seven min, a sufficiently long time to reach a stationary wave turbulence state. The forcing is then stopped at $t=0$, and the temporal decay of the wave amplitude $\eta(t)$ is recorded with a 500 Hz sampling frequency by means of two capacitive probe (C1 and C2) during $15$ minutes, a sufficiently long time to observe the wave damping up to a still state. The experiment is then %automatically
repeated 20 times to improve statistics, and the results are averaged. Accuracy on the wavemaker stopping time is within 2 s. The results are found to be independent of the locations of the probes on the 4 m probe rack. The results reported in this Section do not depend qualitatively on the initial forcing conditions used in table \ref{tabnantes}. 

\begin{figure}
\begin{center}
\includegraphics[scale=0.4]{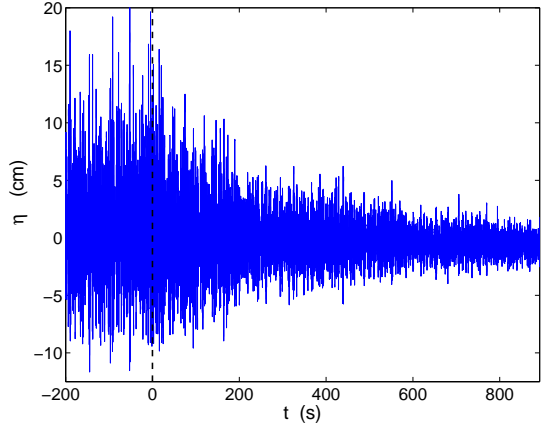} %fig8.eps
\end{center}
\caption{Decay of wave amplitude $\eta(t)$ as a function of time for the reflecting boundary condition. Forcing is working for $t<0$ and is stopped at $t=0$. Initial forcing conditions: FRN forcing with a narrow bandwidth ($f_m=1.15$ Hz, $\Delta f=0.3$ Hz, $\sigma_{\eta}= 4.6$ cm).
} 
    \label{decayfig}
 \end{figure}

\subsection{Temporal decay of the wave amplitude}
The temporal decay of the wave amplitude $\eta(t)$ is shown in figure \ref{decayfig} for a reflecting boundary condition. $t=0$ corresponds to the moment the wavemaker stops. The decay lasts roughly 900~s, including the very slow relaxation of the transverse modes of the tank. Wave energy is dissipated by viscous mechanisms (in bulk, on the free interface, and on the tank sides), and transferred to other scales by nonlinear interactions. For the absorbing boundary condition and for the same initial forcing conditions, the decay is much faster ($\sim 50$ s, roughly corresponding to the propagation time of the last generated wave train) since the beach absorbs most of the wave energy. Thus, we will only report below results on the decay within the closed basin. 

\subsection{Temporal decay of the spectrum}
\begin{figure}
 \begin{center}
\includegraphics[scale=0.4]{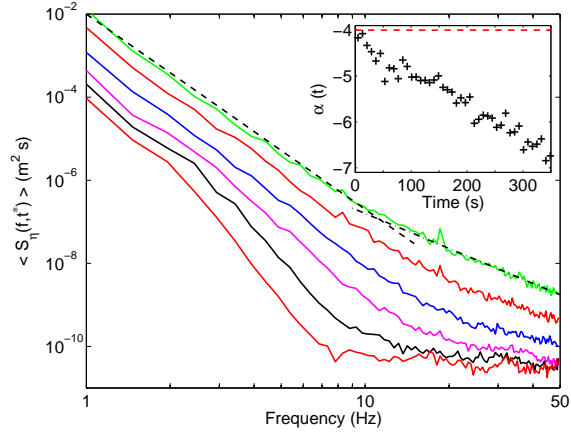}  %Declincompspectra195mm
\caption{Wave spectrum $ \langle S_\eta (f,t^{\ast}) \rangle$ at different times $t^{\ast}$ of the decay. From top to bottom: $t^{\ast} =$ 25, 81, 161, 241, 401, and 641 s. Dashed line is a power law fit $\sim f^{\alpha}$ with $\alpha=-4.7$ in gravity frequency range ($1 \leq f \leq 10$ Hz). Dot-dashed line corresponds to the stationary capillary wave turbulence prediction of $f^{-17/6}$. The three top curves have been shifted vertically for clarity by a factor 10, 5, and 2, respectively. Inset: Gravity exponent $\alpha$ as function of time. Closed basin. Same initial forcing conditions as in figure \ref{decayfig}.} 
    \label{Declincompspectra}
       \end{center}
 \end{figure}

To analyze the different steps of the decay of $\eta(t)$, the time-frequency wave amplitude spectrum $S_\eta (f,t)$ is computed by means of a spectrogram analysis (MATLAB function), for each experiment on short temporal windows $[t,t+\delta t]$ with $\delta t=8$ s, and $0\leq t\leq 800$~s. $S_\eta (f,t)$ is then averaged first over the two probe signals, and then over twenty different realizations leading to the averaged spectrum $ \langle S_\eta (f,t) \rangle$, $\langle \cdot \rangle$ denoting ensemble average. Figure \ref{Declincompspectra} shows $\langle S_{\eta}(f,t^{\ast})\rangle $ as a function of the frequency at different decay times $t^{\ast}$. At the beginning of the decay (top curve), the spectrum displays a frequency-power law $\sim f^{\alpha}$ in the gravity frequency range ($1 \leq f \leq 10$ Hz) with an exponent $\alpha$ close to its value in the stationary regime. When $t^{\ast}$ increases, the power-law spectrum becomes progressively steeper, with $\alpha$ decreasing over time as shown in the inset of figure \ref{Declincompspectra}. No self-similar decay is thus observed in the gravity regime. On the contrary, in the capillary frequency range,  when measurements are not too noisy, it can be observed that the shape of the power law spectrum does not depend significantly on the decay time. This last result is compatible with the self-similar decay of capillary wave turbulence observed previously in a small container \citep{Deike2012}.

 \begin{figure}
 \begin{center}
\includegraphics[scale=0.4]{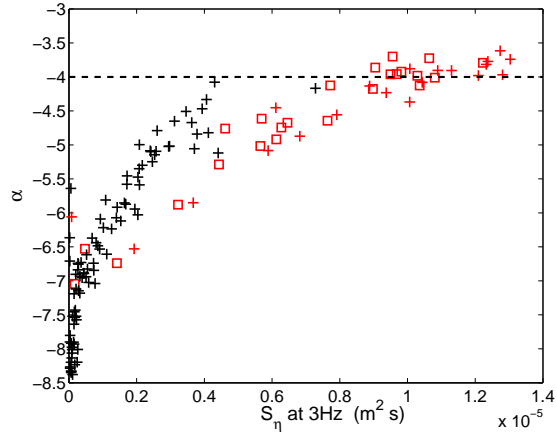}  %fig13.eps
\caption{Frequency-power law exponent, $\alpha$, of the gravity wave spectrum $S_{\eta} \sim f^{\alpha}$ as a function of the value of the spectrum amplitude at 3 Hz, $S_{\eta}(3{\ \rm Hz})$. Decaying regime (black) or stationary regime [red (light grey), same data as in figure~\ref{exposant}]. Reflecting boundary condition. Dashed line corresponds to the prediction of weak turbulence theory $\alpha=-4$. FRN forcing with a broad ($\square$) or narrow ($+$) bandwidth. } 
    \label{pgravvsPSD3Hzdeclin}
       \end{center}
 \end{figure}

\subsection{Frequency-power law exponent of the gravity spectrum during the decay} 
The frequency-power law exponent $\alpha$ of the spectrum is estimated within the gravity frequency range, at each instant of the decay (see inset of figure \ref{Declincompspectra}) and is displayed in figure \ref{pgravvsPSD3Hzdeclin} (black symbols) as a function of $S_{\eta}(3{\ \rm Hz})$, the value of the spectrum amplitude at 3 Hz. Just after the forcing is stopped [corresponding to the highest value of $S_{\eta}(3{\ \rm Hz})$], the exponent of the power spectrum is similar to that observed in the stationary regime, close to $-4$. During the decay, the spectrum decreases in amplitude [smaller values of $S_{\eta}(3{\ \rm Hz})$] and is steeper (see figure \ref{Declincompspectra}). This leads to the exponent $\alpha$ strongly depending on $S_{\eta}(3{\ \rm Hz})$ as shown in figure \ref{pgravvsPSD3Hzdeclin}. The values of $\alpha$ in the decaying regime are then compared in figure \ref{pgravvsPSD3Hzdeclin} with those obtained in the stationary regime (red light grey symbols) of Sect. \ref{SSpectre}, both being performed in the closed basin. We observe that $\alpha$ increases with the spectrum amplitude both in the stationary and decaying regimes with the same trend. As a first approximation, this means that decaying wave turbulence can be seen, at each time point of the decay, as  wave turbulence in a stationary regime but with the corresponding decreasing wave energy. This feature has also been observed for capillary wave turbulence decay \citep{Deike2012,Kolmakov04}. Note that the data for the decaying regime in figure \ref{pgravvsPSD3Hzdeclin} are more scattered than in the stationary regime, as the nonstationary spectra involve less statistics and thus a lower signal to noise ratio.

 \begin{figure}
 \begin{center}
\includegraphics[scale=0.4]{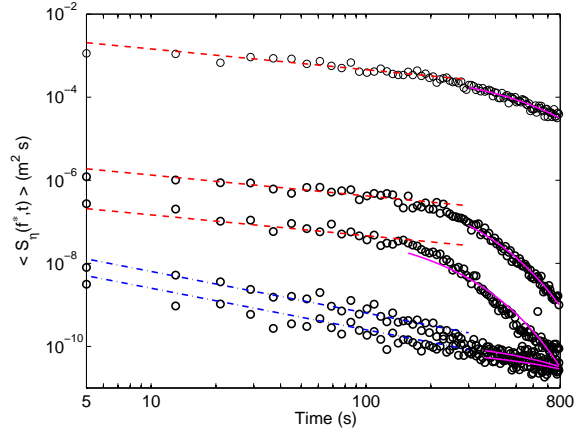}  % fig12.eps Declintempspectra195mm
\caption{Temporal evolution of the wave spectrum $ \langle S_\eta (f^{\ast},t) \rangle$ for Fourier components $f^{\ast}=$1.2, 4.4, 8.3, 18.1, and 28.8 Hz (from top to bottom). %Ecart entre deux points = 8s. 
Red dashed lines: $t^{-1/2}$ law predicted for four-wave interactions (gravity). Blue dot-dashed lines: $t^{-1}$ law predicted for three-wave interactions (capillary). Solid lines: exponential decay $e^{-t/\tau_{d}(f^{\ast})}$ as expected for a viscous damping, with $\tau_{d}(f^{\ast})$ the damping time. Closed basin.} 
    \label{Declintempspectra}
       \end{center}
 \end{figure}
 
 \subsection{Temporal decay of the spectrum and energy Fourier modes}\label{Fouriermodes}
Figure \ref{Declintempspectra} shows the temporal evolution of $\langle S_{\eta}(f^{\ast},t) \rangle$ for different Fourier components $f^{\ast}$. In the first stage of the decay ($t < 200$ s), the Fourier modes in the gravity frequency range are observed to decrease over time as $t^{-1/2}$ (see dashed lines) as predicted for a nonlinear wave decay involving four-wave interactions \citep{NazarenkoAdvance2013}. This confirms with more accuracy the experimental $t^{-1/2}$ scaling found by \cite{Nazarenkojetp2013}. In the capillary frequency range, each Fourier component of the spectrum decreases over time as $t^{-1}$ as expected for three-wave interactions \citep{Falkovich1995} (see dot-dashed lines). This first stage of the decay is thus related to nonlinear mechanisms. For longer decay times ($t > 200$ s), the Fourier modes decay roughly exponentially with time as $e^{-t/\tau_{d}(f^{\ast})}$, as expected for a linear viscous dissipation. Viscous dissipation could arise from surface boundary layers on the bottom and side walls as well as on the free surface due mostly to surface contamination \citep{Lamb1932,VanDorn,Miles1967}. The viscous damping time $\tau_{d}(f^{\ast})$ is fitted empirically and is found to decrease with the Fourier mode frequency from 300 to 100 s typically.  This second stage of the decay is thus driven by the viscous decay of the waves.

Now, let $E_f(t)$ be the wave energy of the Fourier mode at frequency $f$ at time $t$. At $t=0$, the forcing is stopped and the decaying wave energy can be modeled by 
\begin{equation}
\frac{d E_{f}(t)}{dt}= - a_1 E_f(t) - a_2 E_{f}^2(t) - a_3 E_{f}^3(t)\ {\rm ,}
\end{equation} with $a_1$, $a_2$ and $a_3$ taking positive values depending on the frequency $f$. The first term on the right-hand side corresponds to a usual viscous linear dissipation, the second and third term modeling nonlinear dissipation from three-wave and four-wave nonlinear interactions, respectively. These nonlinear dissipations result from the difference at a fixed frequency between the in-flux from low frequencies and the out-flux towards high frequencies. We solve this equation by considering only one non-zero dissipation coefficient $a_1$, $a_2$ or $a_3$ in order to compare the analytical solutions and the experiment results with a unique fit parameter. The linear case leads to $d E_{f}(t)/dt= - a_1 E_{f}(t)$, and thus the wave energy of the Fourier mode decays exponentially in time as
\begin{equation}
E_{f}(t)=E_{f}(0)\exp\left[- t/\tau_d\right]\ {\rm ,}
\label{visc}
\end{equation}
with $1/\tau_d=a_1$ the linear dissipative time scale, and $E_{f}(0)$ the energy when the forcing is stopped. For a quadratic nonlinearity (three-wave interaction such as for capillary waves),  $d E_f(t)/dt= - a_2 {E_f(t)}^2$, and thus 
\begin{equation}E_{f}(t)=E_{f}(0)\left[1+t/\tau^c_{nl}\right]^{-1}\ {\rm ,}
\label{cap}
\end{equation}
with $1/\tau^c_{nl}=a_2 E_{f}(0)$ the nonlinear decay time of capillary waves. For $t \gg \tau^c_{nl}$, this becomes $E_f (t) \sim t^{-1}$. Finally, for a cubic nonlinearity (four-wave interaction such as for gravity waves), one obtains
\begin{equation}
E_{f}(t)=E_{f}(0)\left[1+2t/\tau^g_{nl}\right]^{-1/2}\ {\rm ,}
\label{grav}
\end{equation}
with $1/\tau^g_{nl}=a_3 {E_{f}(0)}^2$  the nonlinear decay time of gravity waves. For $t \gg \tau^g_{nl}$, it follows that $E_f (t) \sim t^{-1/2}$. Note that $\tau_d$, $\tau^c_{nl}$, and $\tau^g_{nl}$ depend on the scale $f$.

The temporal decay of the wave energy $E_{f^*}(t)$ at frequency $f^*$ is related to the power spectrum of wave height at the same component, ${S}_\eta(f^*,t)$, by
\begin{equation}
E_{f^*}(t)=g{S}_\eta (f^*,t)+\frac{\gamma}{\rho}k^2(f^*){S}_\eta (f^*,t) \ {\rm ,}
\label{energiespectre}
\end{equation}
with $k(f)$ given by the dispersion relation of linear gravity-capillary waves. %Equation (\ref{energiespectre}) comes from the expression of the energy of a linear gravity-capillary wave, per unit surface and fluid density, averaged over one period, $E \approx g \eta^2 + \frac{\gamma}{\rho}(\nabla\eta)^2$, and from the definition of the power spectrum normalization $\int_0^{\infty} S_{\eta}(f)df=\sigma_{\eta}^2$ with $\sigma_{\eta}$ the rms value of the wave height $\eta(t)$. 

The temporal decay of the wave energy $E_{f^*}(t)$ at each frequency $f^*$ is thus inferred experimentally from that of the wave spectra, ${S}_\eta(f^*,t)$ -- see figure \ref{Declintempspectra}, by using Eq.\ (\ref{energiespectre}). For a fixed $f^*$ in the gravity range ($0.5\leq f^* \leq 10$ Hz), Eq.\ (\ref{grav}) is found to be a good fit for $E_{f^*}(t)$ over short time periods ($0\leq t \leq 100$ s) leading  to an experimental estimate of $\tau^g_{nl}(f^*)$, $E_{f}(0)$ being given by the value of $\sigma_{\eta}$ in the stationary regime ($t \leq0$). Similarly, for a fixed $f^*$ in the capillary regime ($10 < f^* \leq 50$ Hz), Eq.\ (\ref{cap}) is a good fit for $E_{f^*}(t)$ for small $t$, leading to an estimate of $\tau^c_{nl}(f^*)$. For long times ($t > 200$ s), $E_{f^*}(t)$ is found to decay exponentially in both regimes as in Eqs. (\ref{visc}), thus leading  to an estimate $\tau_d(f^*)$. Finally, reiterating these fits for various $f^*$ gives the frequency dependence of time scales $\tau^g_{nl}$, $\tau^c_{nl}$, and $\tau_d$.

 \begin{figure}
 \begin{center}
\includegraphics[scale=0.4]{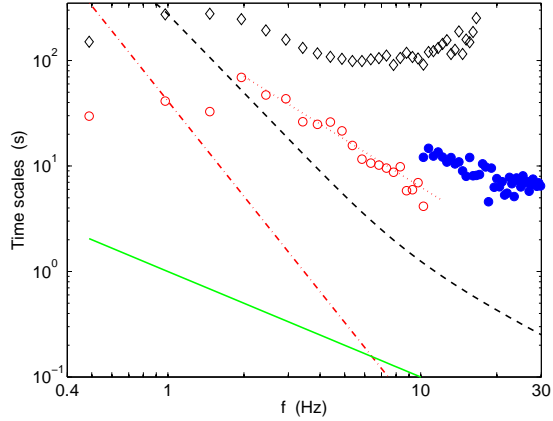}  %Tau.eps
\caption{
Typical time scales as a function of the frequency, $f$. Dissipative linear time scale $\tau_d$ ($\lozenge$). Nonlinear decay times of gravity $\tau^g_{nl}$ ($\circ$), and capillary $\tau^c_{nl}$ ($\bullet$) regimes. $\tau_d$, $\tau^g_{nl}$, and $\tau^c_{nl}$ are inferred from fits of $E(f,t)$ [data of figure \ref{Declintempspectra} using Eq.\ (\ref{energiespectre})] using Eqs. \ (\ref{visc}), (\ref{grav}) and (\ref{cap}), respectively. Solid line: Linear time scale $\tau_l= f^{-1}$. Red dot-dashed line: Theoretical nonlinear interaction time scale of gravity $\tau^g_{4w}\equiv c\epsilon^{-2/3}g^2  f^{-3}$ (see text). Red dotted line: best fit $\sim f^{-3/2}$. Black dashed line: theoretical dissipation time scale $\tau_d^{theo}$ from Eq.\ (\ref{eqdiss}) with $\nu=10^{-6}$ m$^2$/s. Same initial forcing conditions as in figure \ref{decayfig}. Closed basin.
} 
    \label{figtimescale}
       \end{center}
 \end{figure}

\subsection{Time scale separations}
Let us now consider the typical time scales involved in our experiment. Weak turbulence theory assumes a time scale separation $\tau_l(f) \ll \tau_{nl}(f) \ll \tau_d(f)$, between the linear propagation time, $\tau_l$, the nonlinear interaction time, $\tau_{nl}$, and the dissipation time, $\tau_d$. %To our knowledge, such a time scale separation has not been tested experimentally for gravity wave turbulence. 
To our knowledge, such a time scale separation has been  tested experimentally in only two different wave turbulence systems \citep{MiquelPRE2011,Deike2013}, but has never been investigated experimentally for gravity wave turbulence. The linear propagation time is $\tau_l=1/f$, whereas $\tau_d(f)$ and $\tau_{nl}(f)$ are inferred from freely decaying experiments using the results of Sect.\ \ref{Fouriermodes}. These time scales are displayed in  figure \ref{figtimescale}. The dissipative (viscous) linear time scale $\tau_d(f)$ is found to be of the order of 100 s and varies smoothly by a factor of three within the gravity and capillary frequency ranges. For comparison, a theoretical viscous decay time assuming  dissipation due to a viscous surface boundary layer with an inextensible film \citep{Lamb1932,VanDorn,Miles1967,Deike2012} reads
\begin{equation}
 \tau_d^{theo}=\frac{2\sqrt{2}}{k(\omega)\sqrt{\omega \nu}} {\ \rm ,}
 \label{eqdiss}
 \end{equation}
 with $\nu$ the kinematic viscosity of water, and $k(\omega)$ given by the gravity-capillary dispersion relation. This dissipation comes from the presence of surfactants/contaminants at the interface that leads to an inextensible surface where fluid tangential velocity should be cancelled at the interface. This type of dissipation is known to strongly affect the stability of large scale gravity waves in the ocean \citep{HendersonJGR2013}. For all frequencies, $\tau_d$ is found to be much larger than $\tau_d^{theo}$ except at the forcing frequencies $\sim 1$ Hz where the two curves intersect. This observation, and the fact that $\tau_d(f)$ varies smoothly compared to  $\tau_d^{theo}(f)$, mean that the decay of a largest scale mode (near the forcing scale) transfers energy continuously in time towards smaller scales. Thus, the decay of all Fourier modes is driven by the viscous decay of a large scale mode as it has been observed in small container experiments \citep{Deike2012}. Consequently, the estimated nonlinear decay time scales $\tau^g_{nl}(f)$ and $\tau^c_{nl}(f)$ include a contribution due to the cumulative energy transfer from this large scale mode, in addition to that from nonlinear wave interactions. Indeed, $\tau^g_{nl}(f)$ is found to roughly decrease as $f^{-3/2}$ in the gravity inertial range, whereas the four-wave nonlinear interaction time scale reads dimensionally $\tau^g_{4w}\equiv c\epsilon^{-2/3}g^2  f^{-3}$ \citep{Connaughton03,Newell2011} with $\epsilon$ the mean energy flux as estimated in Sect. \ref{KZConstant}, and $c$ a non dimensional constant. $c$ is then adjusted to have $\tau^g_{4w}=\tau^g_{nl}\simeq 40$ s at the forcing frequency $f=1$ Hz. This leads to $\tau^g_{nl}(f)\gg \tau^g_{4w}(f)$ for $f > 1$ Hz, as displayed in figure \ref{figtimescale}. More interestingly, we observe that the scale separation $\tau_l(f) \ll \tau^g_{4w}(f) \ll \tau_d(f)$ is satisfied but in a quite narrow frequency band ($1 < f < 6$ Hz) despite the use of a large basin. Note that a similar analysis can be experimentally performed for the capillary regime (see \citet{DeikeDNS} for direct numerical simulations).
 
Thus, non-stationary experiments make it possible to estimate for the first time the dissipative and nonlinear time scales in gravity wave turbulence at all scales of the cascade by extrapolating their values from that of the forcing scale. We show that an important part of this nonlinear time comes from the cumulative energy transfer from a large scale mode, and thus appears as an upper limit of the four-wave nonlinear interaction time scale of weak turbulence. This large scale mode thus plays a crucial role in gravity wave turbulence in large basins.

\subsection{Estimations of the mean energy flux and Kolmogorov constant}\label{KZConstant}
The mean energy flux cascading from large scales to small scales is a key quantity in hydrodynamics turbulence \citep{Pope}. In wave turbulence, one way to estimate the mean energy flux $\epsilon$ is to measure the wave energy decay rate after switching off the wave maker \citep{Denissenko2007,NazarenkoJFM2010}. This method gives a good estimate of the mean energy flux, provided large scale dissipation is negligible (otherwise the large scale waves lose most of their energy through large scale dissipation rather than by transferring energy to smaller scales). Here, the estimate of $\epsilon$ is obtained just at the beginning of the energy decay, thus avoiding this bias. Assuming no forcing and dissipation, the power budget then reads $dE(t)/dt=-\epsilon$ where $E(t)$ is the wave energy per unit surface and fluid density at time $t$, and $\epsilon$ the mean energy flux per unit surface and density. The energy of linear gravity waves (neglecting capillary waves) averaged over a small time lag reads $E(t)=g\sigma_{\eta}^2(t)$ where $g$ the acceleration due to gravity. Combining both expressions then leads to an estimation of the mean energy flux in the stationary regime ($t\leq 0$)

\begin{equation}
\epsilon=-g\frac{d\sigma_{\eta}^2(t)}{dt}{\Big |}_{t=0} {\rm \ .}
\label{epsilon}
\end{equation}

\begin{figure}
 \begin{center}
\includegraphics[scale=0.4]{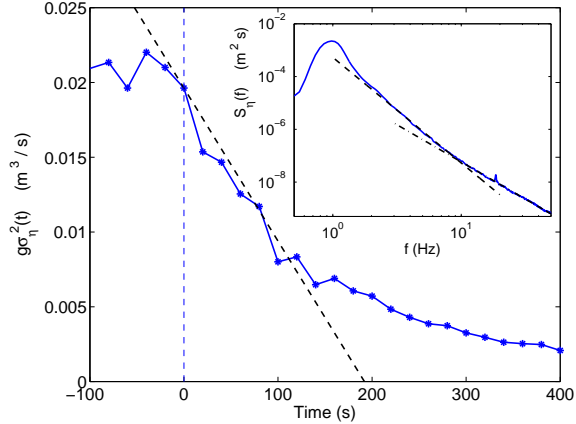}  %Ckz.eps
\caption{Temporal evolution of the gravity wave energy per unit surface and density. The wavemaker is stopped at $t=0$. Dashed line: tangent at $t=0$ of slope $\epsilon=1\times 10^{-4}$ $($m$/$s$)^{3}$ - see Eq.\ (\ref{epsilon}). Each value of $\sigma_{\eta}$ is averaged over 20 s and averaged over 19 runs to have a good statistical convergence. Same initial forcing conditions as in figure \ref{decayfig}. Closed basin. Inset: Wave power spectrum in the stationary regime for the same forcing conditions. Dashed line: Theoretical weak turbulence spectrum $C\epsilon^{1/3} g \omega^{-4}$ for gravity waves with $C=1.8$ and $\epsilon=1\times 10^{-4}$ $($m$/$s$)^{3}$. Dot-dashed line: $f^{-17/6}$ power-law fit. Closed basin.} 
    \label{figenergyflux}
       \end{center}
 \end{figure}

Figure \ref{figenergyflux} shows the temporal evolution of $E(t)$ after switching off the wave maker at $t=0$. The tangent to the curve at $t=0$ then gives $\epsilon=100 \pm 30$ $($cm$/$s$)^{3}$. Note that this value is much smaller than the critical flux  $(\gamma g / \rho)^{3/4} \approx 2200$ (cm/s)$^3$ corresponding to the breakdown of weak turbulence at the transition between gravity and capillary regimes \citep{Newell92}. Estimated values of $\epsilon$ in our experiments are such that $\epsilon < (\gamma g / \rho)^{3/4}$. Our estimate of $\epsilon$ from the decay of the wave energy is found to increase as expected when the initial wave amplitude increases.

It is now possible, knowing the value of $\epsilon$, to evaluate experimentally the Kolmogorov constant $C$ of Eq. (\ref{gravsptheo}) from the gravity wave spectrum obtained in the stationary regime at sufficiently high  forcing. The inset of figure \ref{figenergyflux} shows such a spectrum displaying  good agreement with the $\omega^{-4}$ power-law scaling expected in the gravity wave turbulence regime and the $\omega^{-17/6}$ scaling expected in the capillary regime. Using the $\omega^{-4}$ fit parameter, the value of $\epsilon$ obtained above, and the expression of the non dimensional Kolmogorov-Zakharov constant \citep{Zakharov1967}
\begin{equation}
C=\frac{S_{\eta}(\omega)\omega^{4}}{\epsilon^{1/3}g}\ {\rm ,}
\end{equation}
 [$S_{\eta}(\omega)$ has dimension $L^2T$ and $\epsilon$ has dimension $L^3/T^3$], one finds a value of the constant $C=1.8\pm 0.2$ of the same order of magnitude as a theoretical value of $2.75$ estimated by \cite{Zakharov2010}. Note that \cite{BadulinNPG05} found a numerical constant value of 0.5. Our study therefore  reports the first experimental estimation of the Kolmogorov-Zakharov constant for gravity wave turbulence, the latter being compatible with a recently obtained theoretical value.

 \section{Conclusion}
We have reported results of experiments on gravity wave turbulence in a large basin. The role of the basin boundary conditions has been tested. To this end, an absorbing sloping beach opposite the wavemaker can be replaced by a reflecting wall. We observe that the wave field properties depend strongly on these boundary conditions. A quasi one-dimensional field of nonlinear waves propagates toward the beach where they are damped whereas a more multidirectional wave field is observed with the wall. In both cases, the wave spectrum shows power-law scalings over a two-decade frequency-range (one decade in the gravity range and one in the capillary range). The frequency-power law exponent of the gravity spectrum is found to depend on the nonlinearity level (i.e. forcing strength) with a similar trend in both cases, and up to a value close to $-4$ at sufficiently high  nonlinearity. The physical mechanisms leading to this spectrum at high nonlinearity are likely to be different: mainly due to propagation of coherent structures in the presence of a beach and to interactions between nonlinear waves in the presence of a wall. The observed steepening of the spectrum at low nonlinearity, in both cases, could be explained by the dissipation occurring at all scales of the turbulent cascade (see below), a situation not taken into account so far by weak turbulence theory. Small scale intermittency properties of gravity wave turbulence have then been quantified. We found roughly the same value of the intermittency coefficient in the presence of either a beach or a wall, suggesting the importance of coherent structures in both cases. 

We have also studied the non-stationary regime of gravity wave turbulence during its free decay. No self-similar decay is observed in the gravity regime (the frequency power-law exponent of the instantaneous spectrum being dependent on time). We also show that the spectrum Fourier mode amplitudes decay first as a time power law due to nonlinear mechanisms, and then exponentially due to linear viscous damping. A new estimate  of the mean energy flux is obtained from the initial decay of wave energy.  The Kolmogorov-Zakharov constant is then evaluated for the first time at high nonlinearity, and found to be compatible with a theoretical value estimated by \cite{Zakharov2010}. We have also inferred the linear, nonlinear, and dissipative time scales at all scales of the cascade. The time scale separation highlights the important role of a large scale Fourier mode (near the forcing scale). Such a large scale mode probably generates non-local interactions that are not yet taken into account  in weak turbulence theory. 

Finally, we have found that viscous dissipation occurs at all scales of the cascade, contrary to theoretical hypothesis, and thus induces an ill-defined inertial range between forcing and dissipation. The relative importance of dissipation with respect to nonlinear interactions may explain the observed steepening of the gravity spectrum at low nonlinearity. Indeed, a similar phenomenon has previously been reported both experimentally, in studies of wave turbulence on a metallic plate \citep{HumbertEPL2013,MiquelPRE2014}, and of capillary wave turbulence \citep{Deike2014} when increasing dissipation (e.g. adding dampers on the plate, or working with high enough viscosity fluids), and numerically when reducing the nonlinear interactions \citep{Pan2014}. Here also, the ratio between dissipation and nonlinearity has to be small enough at all scales to reach a wave turbulence regime. Further theoretical developments introducing realistic empirical dissipating terms in the kinetic equation (as tested numerically by \citet{ZakharovPRL07,WiseGroup2007} and references therein) would therefore be of primary interest for improving understanding of gravity wave turbulence in large basins. Although these experiments cannot reproduce real ocean conditions, they could help to understand and to model fully developed and self-similar regimes of swell, which result as an equilibrium between wind input, nonlinear wave-interactions and dissipation \citep{Gagnaire11,Korotkevich08,Phillips85}.

\begin{acknowledgments}
This work was supported by ANR Turbulon 12-BS04-0005. We thank S. Auma{\^{i}}tre, S. Fauve and F. P\'etr\'elis for fruitful discussions. We thank L. Davoust, S. Lambert, C. Laroche and J. Servais for their technical help. We also thank S. Nazarenko and S. Lukaschuk for sending us their data \citep[][figure 6 left]{NazarenkoJFM2010}.
\end{acknowledgments}

\bibliographystyle{jfm}
\bibliography{Turbulonbib}

\end{document}